\documentclass{aa}  

\usepackage{graphicx}
\usepackage{float}
\usepackage{txfonts}
\usepackage[T1]{fontenc}
\usepackage{courier}
\usepackage{comment}
\usepackage{multirow}
\usepackage{hyperref}
\hypersetup{colorlinks=true, citecolor=blue, linkcolor=blue, filecolor=magenta, urlcolor=cyan, pdftitle={Overleaf Example}, pdfpagemode=FullScreen}
\usepackage[export]{adjustbox}

\begin{document} 

   \title{Chemical decoding of kinematic substructures in the Galactic halo} 

   \author{A. Mori\inst{1,2}
          \and P. Di Matteo \inst{3} 
          \and S. Salvadori \inst{1,2}
          \and M. Mondelin \inst{4}
          \and S. Khoperskov \inst{5}
          \and M. Haywood \inst{3}
          \and A. Mastrobuono-Battisti \inst{6,7,8}
          }

   \institute{Dipartimento di Fisica e Astronomia, Università di Firenze, Via G. Sansone 1, 50019 Sesto Fiorentino, Firenze, Italy\\
         \email{alice.mori@unifi.it}
         \and INAF – Osservatorio Astrofisico di Arcetri, Largo Enrico Fermi 5, 50125 Firenze, Italy
         \and LIRA, Observatoire de Paris, Universit\'e PSL, Sorbonne Universit\'e, Universit\'e Paris Cit\'e, CY Cergy Paris Universit\'e, CNRS, 92190 Meudon, France
         \and CEA, AIM, Université Paris-Saclay, 91191 Gif-sur-Yvette, France
         \and Leibniz Institut für Astrophysik Potsdam (AIP), An der Sternwarte 16, 14482 Potsdam, Germany
         \and Dipartimento di Fisica e Astronomia “Galileo Galilei”, Università di Padova, Vicolo dell’Osservatorio 3, 35122 Padova, Italy 
         \and Dipartimento di Tecnica e Gestione dei Sistemi Industriali, Università di Padova, Stradella S. Nicola 3, I-36100 Vicenza, Italy
         \and INAF – Osservatorio Astronomico di Padova, Vicolo dell’Osservatorio 5, Padova, I-35122, Italy
              }

   \date{Accepted by A\&A
   }
 
  \abstract
   {In the hierarchical mass assembly framework, the accretion history of the Milky Way is crucial to understanding its evolution. Previous studies have shown that integrals of motion are not strictly conserved during massive mergers, leading to the broad redistribution of accreted stars across dynamical spaces, such as energy-angular momentum ($E-L_z$). Additionally, part of the in situ disc component itself becomes kinematically heated and acquires halo-like orbits as a result of these interactions. Consequently, even for minor mergers, which experience weaker dynamical friction and are then supposed to preserve a higher degree of phase-space coherence, we expect their kinematic-defined samples to be anyway contaminated by both the massive merger(s) and the disc stars, thereby hiding the presence of different populations within their samples.}
   {This study aims at quantifying these contamination effects in known accreted halo substructures. As they are defined by their kinematics, we aim at cleaning their samples analysing their chemical properties only, to uncover their specific abundance patterns.} 
   {We applied the kinematic selection criteria for the halo substructures to the Gaia EDR3 and APOGEE DR17 data. Then we adopted a Gaussian Mixture Model approach to chemically compare different substructures on a star-by-star basis, taking into account several abundances (Fe, Mg, Si, Ca, Mn, Al, and C). This method incorporates the analysis of various elements that probe different nucleosynthetic channels, providing a percentage of global chemical compatibility. 
   }
   {We argue that the chemical properties of Sequoia point towards a shared origin with GSE. Heracles, Thamnos and the Helmi Stream all likely comprise GSE and heated disc stars in a significant amount. 
   Besides these two populations, we identified stars with chemical abundances and orbital properties compatible with the Sagittarius stream within the Helmi Stream and with $\omega$ Cen within Thamnos. Finally, GSE itself is contaminated by the Sagittarius Stream.}
   {Halo stars chemically compatible with GSE are spread throughout the $E-L_z$ space and considerably contaminate every halo substructure studied in this work. None of these substructures appears to be a unique population of stars with its own origin. In addition to GSE, they all appear to be mixtures of stars chemically compatible either with the metal-poor disc, Sagittarius, $\omega$ Cen, or with a combination of them.}


   \titlerunning{Chemical decoding of kinematic substructures in the Galactic halo}
   \maketitle

\section{Introduction}
In the hierarchical mass assembly framework predicted by the standard cosmological model, the accretion history of the Milky Way is key to unravelling its formation and evolution. The identification of ex situ stellar populations in the stellar halo has been carried out over the last decades on the basis of chemo-dynamical arguments \citep[see][for recent reviews on Galactic Archaeology with 6D phase-space plus chemical information from Gaia and its complementary spectroscopic surveys]{helmi20, deasonbelokurov24}. In particular, accreted substructures have been discovered assuming that they exhibit small spreads in integrals of motion, such as energy and angular momentum \citep[e.g.,][]{helmiwhite99, helmidezeeuw2000, knebe2005, brown2005, mcmillanbinney08, gomez2010}, along with matching chemical abundance properties, generally such as metallicity distribution functions or sequences in the $\alpha$-metallicity plane \citep[e.g.,][]{freemanblandhawthorn02, venn2004, nissenschuster2010, hawkins15, haywood2018, naidu2020}.

These assumptions have led to the reconstruction of several merger events experienced by the Milky Way over cosmic time. Among the plethora of discovered substructures we report the most significant ones in a non-exhaustive way: the last massive merger of the Galaxy with \textit{Gaia} Sausage/Enceladus \citep[hereafter GSE,][]{belokurov2018, haywood2018, helmi2018, vincenzo19, mackereth19, gallart19, feuillet20}, retrograde debris such as Sequoia \citep[][]{myeong2019, naidu2020, matsuno22a} and Thamnos \citep[][]{koppelman19b, ruiz-lara22}, prograde ones as the Helmi Stream \citep[][]{helmi1999, koppelman19a}, and the most bound one which is Kraken/Koala/Heracles \citep[][]{kruijssen19, kruijssen20, forbes20, horta21}. The interpretation of the latter signature is still debated, as also associated with the in-situ component Aurora or proto-Milky Way \citep[e.g.,][]{belokurovkravtsov22}.
Moreover, we can appreciate the ongoing merger with the Sagittarius dwarf spheroidal galaxy as traced by its stream and by the galaxy itself \citep[][]{ibata1994, majewski2003} and with the Magellanic Clouds \citep[][]{putman98}.

However, more recent studies have shown that the integrals of motion are not necessarily conserved when massive merger events are concerned, because of dynamical friction and time-dependent, non-axisymmetric potentials. In fact, the accreted stars spread across the dynamical spaces, thus not being clumped in a single region and possibly producing several overdensities \citep[][]{jean-baptiste2017, grand19, koppelman20, amarante22, khoperskov23a, mori24, thomas25}. Taking into account more than one satellite, the simulations show that the merger debris overlap with each other in the spaces mentioned above \citep[][]{jean-baptiste2017}. Additionally, part of the in situ disc component itself becomes kinematically heated and acquires halo-like orbits as a result of the merger, thus overlapping with accreted populations and forming its own substructures \citep[][]{dimatteo2019, belokurov20}. As a consequence of all these dynamical considerations, even if minor mergers are supposed to preserve a higher degree of phase-space coherence because of the lower dynamical friction effect, we expect their kinematic-defined samples to be anyway contaminated by both the massive merger(s) and the heated disc stars, thereby hiding different populations. 

More specifically, we know that the Galaxy has experienced at least one massive merger (i.e., with GSE), which likely has a broader distribution in the energy-angular momentum space with respect to the sharp kinematic cuts by which it is defined in the literature \citep[e.g., in ][]{horta23, koppelman19b, massari2019}. It could have then produced clumps in other regions of the space that may be associated to independent mergers, it could have contaminated the neighbouring ones produced by different accretions, and finally it may have triggered the kinematic heating of disc stars, which could contaminate the accreted clumps as well \citep[e.g.,][]{jean-baptiste2017, koppelman20, skuladottir25, berni26}. To complicate the picture, a number of studies have also shown that accreted debris can generate metallicity gradients in the energy angular momentum space \citep[][]{amarante22, khoperskov23c, mori24}. If these accretion events are sufficiently massive, they can, in addition, produce clumps featuring different metallicity distribution functions (with different shapes and mean values), despite being related to the same merger event \citep[][]{mori24}.

It is clear that the overall outcome of these premises is a very complex picture to disentangle. Today we can observe various substructures in the Milky Way's stellar halo, defined according to a relative difference in kinematics and chemical properties. However, is there really a way to discriminate between different accretion debris and to assess whether two or more of them are related to each other? This is the fundamental question guiding this work. In particular, we aim at establishing whether putative present-day substructures are connected to the GSE event or to heated disc stars. Once both sources of contamination have been quantified, we may finally be able to identify genuine halo substructures and uncover their specific properties.

As the chemical abundances of a group of stars can highlight different star formation histories and thus distinct birth environments \citep[][]{freemanblandhawthorn02, bland-hawthorngerhard2016}, we can exploit this information for our purposes. Previous studies have attempted to characterise halo substructures through their chemical properties \citep[][]{monty20, hasselquist21, feuillet21, naidu22, matsuno22a, matsuno22b, buder22, horta23, ceccarelli24, ceccarelli25}. However, throughout these works, the contamination of the samples has never been systematically quantified; there are only a few cases which analyse the contamination of GSE in either Sequoia or Thamnos \citep[][]{matsuno22a, dodd25, ceccarelli25}. Chemical trends of different abundances have been derived and compared to each other for various substructures, assuming that they were pure. The issue arises in the case in which a sample is composed of multiple populations: the resulting chemical pattern of the substructure would then appear as a mixture of the two, affecting the median values. 
When compared to the stellar population of another substructure, this mixed sample could then appear as distinct, even though possibly comprising in a significant fraction the population to which it is compared. In order to be able to capture the comprehensive chemical information of the mixture, we would need to take into account the overall distribution of the abundances, which might exhibit the chemical imprints of their multiple components. This kind of analysis can be done using Gaussian Mixture Models (GMM), which allow for the comparison of two samples, considering the whole distribution of different abundances all at the same time and provide a fraction of chemically compatible stars. 

By applying this method to data from \textit{Gaia} and APOGEE, we can perform a revision of the kinematically defined halo substructures known in the literature by means of a comprehensive chemistry-only analysis. The main aim is quantifying how much different halo substructures overlap with each other, GSE and in situ stellar populations in the chemical abundance space. In this way, we will define how large the distribution of GSE in the energy-angular momentum space is, including the chemically compatible stars that have different kinematic properties but are still potentially linked to GSE. Moreover, we will find whether there are any clumps with specific chemical patterns distinct from GSE/disc, and - in case there are - we will characterise their cleaned samples. 

The outline of the paper is described below. The observational data and the selection criteria assumed are detailed in Sect. \ref{sec:data}. The kinematic criteria that define the halo substructures in previous works are briefly illustrated in Sect. \ref{sec:kin}, while their main chemical abundance properties are described in Sect.~\ref{sec:abundances}. The description of the method we adopt to compare the samples, that is the definition of their Gaussian Mixture Models, can be found in Sect. \ref{sec:method}. The results of the chemical compatibility of the halo substructures with respect to the GSE and the disc samples are reported in Sect. \ref{sec:res:gse} and Sect. \ref{sec:res:disc}, respectively. The substructures' samples cleaned from both sources of contamination are finally described in Sect. \ref{sec:res:rest}. We then discuss the results in Sect. \ref{sec:discussion}, and summarise the conclusions in Sect. \ref{sec:conclusions}.

\section{Observational data}
\label{sec:data}

The observational data analysed in this work are cross-matched from the early third data release of the ESA astrometric mission \textit{Gaia} \citep[EDR3; ][]{gaiaedr3} and the latest data release of the SDSS spectroscopic survey APOGEE \citep[DR17; ][]{abdurrouf22}. Moreover, we combined the APOGEE Value Added Catalogue (VAC) \texttt{astroNN} \citep[][]{leungbovy19b}, that provides distances from applying the astroNN deep-learning code to the APOGEE DR17 spectra retrained with Gaia EDR3. This combination allows for full 6D phase-space information for over 700000 Milky Way stars, as it accounts for coordinates, distances (d), proper motions (PM), and radial velocities (RV); along with chemical abundances for over 20 elements. We then make use of the APOGEE VAC of Galactic globular cluster (GC) stars \citep[][]{schiavon24}, in order to remove 7737 GC candidates, identified with the membership probability (\texttt{VB\_prob}), determined via a Gaussian mixture model of the Gaia EDR3 positions and PMs of GC stars \citep[][]{vasilievbaumgardt21, baumgardtvasiliev21}. Stars with \texttt{VB\_prob} > 0.8 were excluded. Eventually, we took advantage of the APOGEE samples of Massive Milky Way Satellites \citep{hasselquist21}, which enabled us to also remove stars belonging to the Small and Large Magellanic clouds (5055 in total, 1146 from the SMC and 3909 from the LMC). After having collected our sample, we applied the following selection criteria \cite[as in][]{horta23}: APOGEE atmospheric parameters: $3500 < T_{eff}/K < 5500$ and $log \textit{g} < 3.6$ to select giants, spectral S/N > 70, \texttt{STARFLAG} = 0, astroNN distance accuracy of $d_{\odot,err}/d_{\odot} < 0.2$ ($d_{\odot}\ge0$), APOGEE \texttt{X\_FE\_FLAG} = 0 and no \texttt{nan} values. The final sample then comprises 168859 stars of the Milky Way (hereafter MW).  

To unravel whether stars of different halo substructures were born in distinct environments, thus possibly in different galaxies, we can study their star formation and chemical evolution histories by analysing their chemical abundances. The abundances of the stellar atmospheres are fossil records of the interstellar medium out of which they formed, that had already been polluted by previous generations of stars exploded as supernovae or by stellar winds from AGB stars. Accordingly, different abundances may trace distinct environmental chemical enrichment, due to diverse star formation histories, pointing towards an independent origin. Given this global picture, we would aim at analysing the highest possible number of chemical elements, as they trace different nucleosynthetic channels. In this sense, the more information we have for the abundances, the better we can discriminate between different star formation histories in principle. For this scope, APOGEE is an excellent tool, providing chemical abundances for more than 20 elements. However, not all of their measurements are given with the same precision and free of quality flags. As to fulfil the selection criteria reported above, but at the same time preserve a statistically relevant number of stars in each analysed substructure, we decided to consider the following abundance ratios: [Fe/H], [Mg/Fe], [Si/Fe], [Ca/Fe], [C/Fe], [Al/Fe], and [Mn/Fe]. In this way, we considered the most reliable elements in APOGEE for the various nucleosynthetic categories: Mg for $\alpha$, Al for odd-z and Mn for the iron-peak, driven by the APOGEE's recommendations for giant stars. Thus we obtained high-precision samples (median uncertainties $\le$ 0.02) that still allowed us to probe several nucleosynthetic channels. Beyond analysing iron, mainly produced by type Ia supernovae, we study then $\alpha$-elements (like Mg, Si and Ca), mainly produced instead by type II core-collapse supernovae, a light element like carbon, as to evaluate ancient high-mass star pollution versus later AGB enrichment, an odd-Z element like aluminum to assess the massive star contribution and an iron-peak one that is manganese, to consider the relative impact of type Ia supernovae.

\section{Kinematic substructures}
\label{sec:kin}

\begin{table*}[]
    \caption{Selection criteria for the halo substructures.}
    \label{tab:kin}
    \centering
    \resizebox{\linewidth}{!}{
    \begin{tabular}{ c | c c c c | c | c | c }
        \hline 
        \hline
        Name & \multicolumn{5}{c |}{Dynamical (and chemical) selection criteria} & $N_*$ & Ref \\
        \hline
        \hline
        & $E$ & $L_z$ & $L_{\perp}$ & $e$/$\eta$ & Abu & & \\
        \hline
        GSE (H)          & $-1.6<E<-1.1$  & $-0.5<L_z<0.5$          & -                             & -                 & n &  941 & H23\\
        GSE (K)          & $-1.5<E<-1.1$  & -                       & -                             & $-0.20<\eta<0.13$ & n &  442 & K19a\\
        GSE (M)          & $-1.86<E<-0.9$ & $-0.8<L_z<0.62$         & $L_{\perp}<3.5$               & -                 & n & 3332 & M19\\
        Sequoia (K)      & $-1.35<E<-1$   & $L_z<0$                 & -                             & $-0.65<\eta<-0.4$ & n &   18 & K19a\\
        Sequoia (My)     & $E>-1.5$       & $J_{\phi}/J_{tot}<-0.5$ &  $J_{(J_z-J_R)}/J_{tot}<0.1$  & -                 & n &  107 & My19\\
        Sequoia (M)      & $-1.5<E<-0.7$  & $-3.7<L_z<-0.85$        & -                             & -                 & n &  131 & M19\\
        Helmi Stream (K) & -              & $0.75<L_z<1.7$          & $1.6<L_{\perp}<3.2$           & -                 & n &   54 & K19b\\ 
        Helmi Stream (M) & $E<-1$         & $0.35<L_z<3$            & $1<L_{\perp}<3.2$             & -                 & n &  427 & M19\\
        Thamnos          & $-1.8<E<-1.6$  & $L_z<0$                 & -                             & $e<0.7$           & n &  118 & K19a\\ 
        Heracles         & $-2.6<E<-2$    & -                       & -                             & $e>0.6$           & y &  222 & H21\\
        \hline
    \end{tabular}
    }
    \tablefoot{$L_z$ and $L_{\perp}$ are in units of $10^3$ kpc km s$^{-1}$, $E$ of $10^5$ km$^2$ s$^{-2}$. The dynamical criteria are based on total orbital energy ($E$), angular momentum with respect to the Galactic disc ($L_z$), perpendicular angular momentum ($L_{\perp}$), azimuthal, vertical and radial action ($J_{\phi}$, $J_{z}$, and $J_{R}$, respectively), eccentricity ($e$), and circularity ($\eta$). The column "Abu" states whether the selection comprises chemical abundance criteria: Heracles (Ho21) is the only case, described in the text. In addition, we analysed the Sgr dSph and $\omega$ Cen GC, whose stars were selected directly from Ha21 and S24 catalogues, respectively. The references are the following. K19a: \citet[][]{koppelman19b}, K19b:\citet[][]{koppelman19a}, M19: \citet[][]{massari2019}, My19: \citet[][]{myeong2019}, H21,23:\citet[][]{horta21, horta23}, Ha21:\citet[][]{hasselquist21}, S24:\citet[][]{schiavon24}. } 
\end{table*}

To build the sample of halo substructures that we want to compare from a chemical point of view, we assumed the kinematic definitions summarised in \citet[][]{horta23}. To this aim, we transformed the phase-space data from the ICRS frame into Galactocentric coordinates by means of the \texttt{astropy} Python library, assuming for the Sun a distance from the Galactic centre of $R_0 = 8.178$ kpc, a vertical height from the midplane of $z_0 = 0.02$ kpc, and a velocity $[U_{\odot}, V_{\odot}, W_{\odot}] = [-11.1, 248.0, 8.5]$ km s$^{-1}$. We then obtained the orbital parameters through the \texttt{galpy} code \citep[][]{bovy15, mackereth18}, assuming the \citet[][]{mcmillan17} Galactic potential. 

\begin{figure}
    \centering
    \includegraphics[width=1\columnwidth]{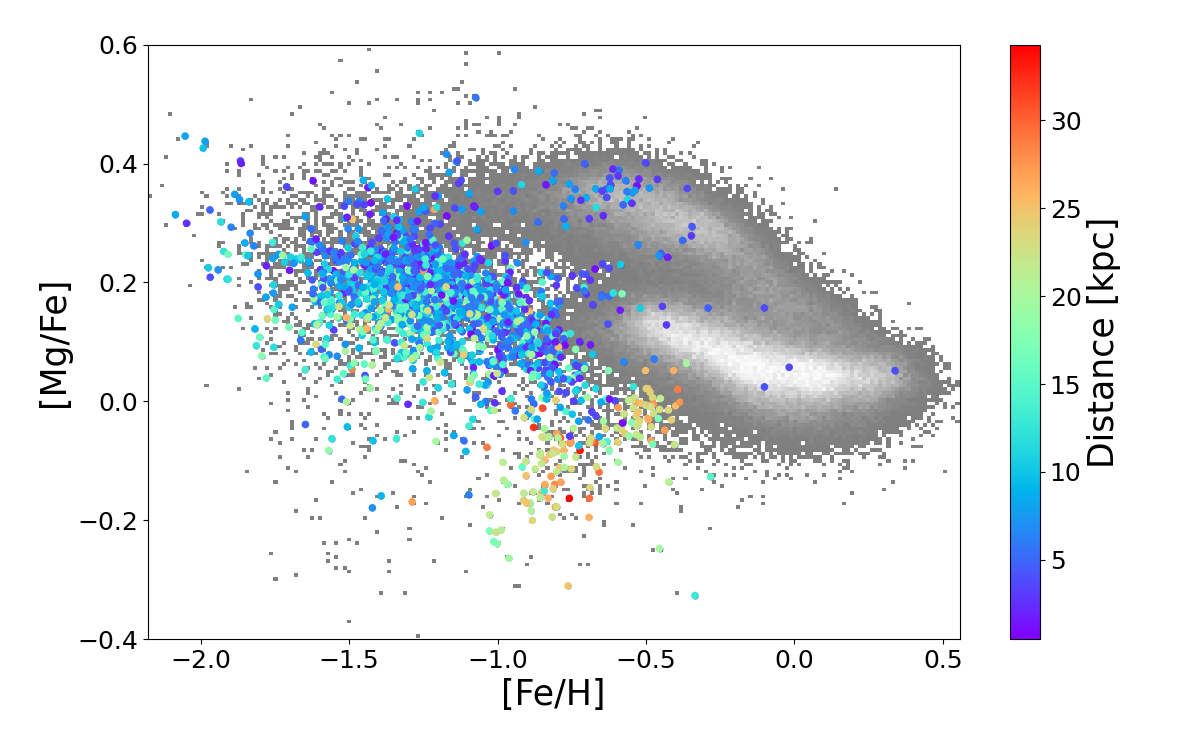}
    \caption{GSE (H) distribution in the [Mg/Fe] vs. [Fe/H] space, colour-coded by the distance from the Sun.}
    \label{fig:gsedist}
\end{figure}

The selection criteria defining the halo substructures are adopted from the literature \citep[also summarised in][]{horta23} and are based on purely kinematic assumptions, where possible. In particular, we analysed the accreted debris of the following merger events (assuming the reported selections): GSE (GSE (H), \citealt{horta23}; GSE (K), \citealt{koppelman19b}; GSE (M), \citealt{massari2019}), Sequoia (Sequoia (K), \citealt{koppelman19b}; Sequoia (My), \citealt{myeong2019}; Sequoia (M), \citealt{massari2019}), Helmi Stream (Helmi Stream (K), \citealt{koppelman19a}; Helmi Stream (M), \citealt{massari2019}), Thamnos \citep[][]{koppelman19b}, Heracles \citep[][]{horta21, horta23}, and Sagittarius \citep[][]{hasselquist21}. 
We note that many of these selections were defined within a smaller volume around the Sun and with different solar parameters in the literature, but we applied them to larger distances to compare with the results of \citealt{horta23}, from which we also adopted the solar parameters.
We also acknowledge that we have considered the selections of \citet{massari2019} for comparison with those of \citet{koppelman19b,koppelman19a, horta23}. These selections were initially defined to associate GCs with known accreted debris, thus they are typically more extended in IOM spaces. 
The criteria, references and number of stars for each sample are reported in Table \ref{tab:kin}. 
We finally note that we excluded all stars with flags for abundance measurements in APOGEE in any of the elements considered in the analysis -- as requested by our method (see Sec. \ref{sec:method}), thus our samples are generally smaller than H23 original ones, who imposed the flag constraint only for the 2 elements considered at each step of the chemical comparison.
We included $\omega$ Cen \citep[][]{schiavon24} in the analyses, as a candidate remnant of an accretion event and possibly the nuclear star cluster of GSE/Sequoia \citep[][]{massari2019, myeong2019, callingham22, limberg22}.
Sgr and $\omega$ Cen are not described in Tab. \ref{tab:kin}, as their stars have been selected directly from the following catalogues: \citet[][]{hasselquist21} and \citet[][]{schiavon24}, respectively.
Note that the definition of the Heracles sample provided by \cite{horta21} and used in this work is based not only on kinematic cuts, but also on the following chemical selection criteria: $\rm{[Al/Fe]} < -0.07 \,\&\, \rm{[Mg/Mn]} \ge 0.25$; $\rm{[Al/Fe]} \ge -0.07 \,\&\, \rm{[Mg/Mn]} \ge 4.25\times \rm{[Al/Fe]} + 0.5475$; $\rm{[Fe/H]} > -1.7$.
Moreover, we defined the in situ disc by circularity\footnote{The circularity is defined as $\eta = L_z/|L_{z,circ}|$.} $\eta > 0.8$, and the metal-poor disc when considering an additional cut in metallicity $\rm{[Fe/H]}\le-0.8$.

In Fig.~\ref{fig:gsedist} we show the distribution of GSE (H) stars in the [Mg/Fe] vs. [Fe/H] space obtained by using the aforementioned criteria and coloured according to their distance from the Sun.
We can notice that beyond the main bulk of GSE stars drawing the typical low-$\alpha$ sequence of accreted dwarf galaxies, there is an orthogonal sequence due to distant stars (red-orange-green points). As the aim of this work is to evaluate the chemical compatibility of the halo substructure with GSE, we need a clean GSE sample, to focus on the chemical properties characteristic of the bulk of its stars. 
For this reason, we further selected GSE stars by imposing a cut in distance (d$\le$ 10 kpc). 
We verified that these distant GSE stars (d $\sim$ 20 kpc) are chemically compatible with Sagittarius and its stream, and some of them (24 out of 1603) are in fact part of the \citet[][]{hasselquist21} sample. We will further comment on the chemical compatibility between GSE and Sagittarius in Section \ref{sec:res:gse}.\\

\begin{figure}
    \centering
    \includegraphics[width=1\columnwidth]{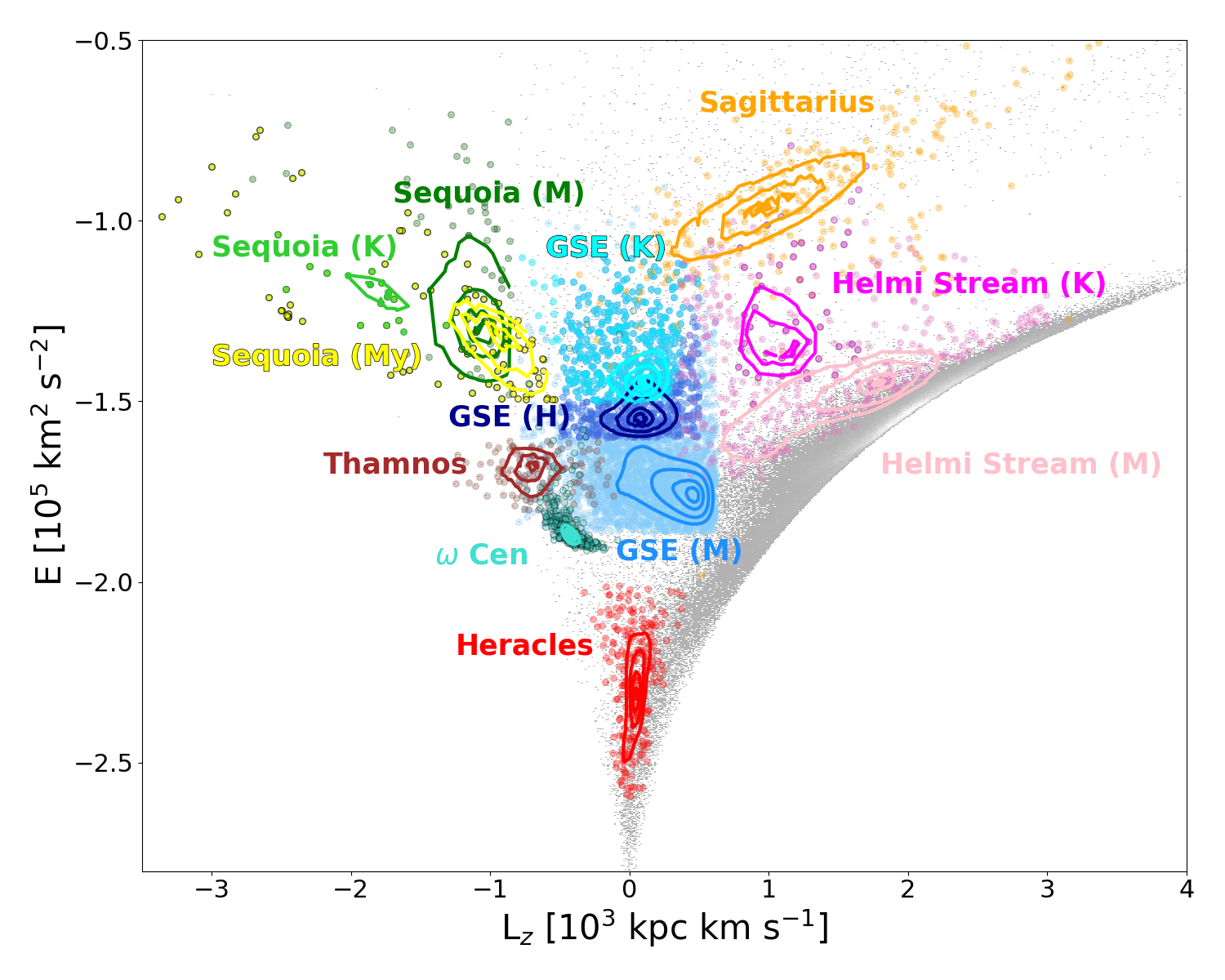}
    \caption{Distribution of the halo substructures in the orbital energy ($E$) versus angular momentum with respect to the Galactic disc ($L_z$) space. Beyond the dynamical definitions in Table \ref{tab:kin}, we report the Sgr dSph and $\omega$ Cen distributions for comparison. 
    The contours represent probability density levels corresponding to 68\%, 85\%, 95\%, 99.7\% of the enclosed data for each substructure, based on Gaussian kernel density estimate, thus showing iso-density levels.
    } 
    \label{fig:elz}
\end{figure}
For a visualisation of these kinematic cuts, in Figure \ref{fig:elz} we show the orbital energy ($E$) versus the angular momentum with respect to the Galactic disc ($L_z$). The different halo substructures are reported in various colours and we can identify them starting from the top (highest energies) to the bottom (lowest energies). At $E \ge -1.7$ we see the bulks of four substructures: the Sagittarius stream and dSph (orange) with mainly prograde orbits $L_z \ge 0.4$ and $E \ge -1.1$; below it ($-1.5 \le E \le -1.1$) and with the same $L_z$ we find the Helmi stream (magenta/pink); then we have the GSE remnant (light blue/dark blue/cyan) with zero angular momentum (radial orbits); on the other hand, the Sequoia debris (dark green/light green/yellow) with retrograde orbits ($L_z \le - 0.4$). In the intermediate energy range, $-2 \le E \le -1.6$, and with retrograde orbits, we see both Thamnos (brown) and $\omega$ Centauri (turquoise). Finally, at very low energy $E \le -2$, there is the Heracles debris (red) with zero angular momentum. 

Thus, by inspecting Figure \ref{fig:elz}, it is clear that the substructures defined by the dynamical criteria overlap partially, although they generally occupy different regions in the $E-L_z$ space. The fact that the substructures reside in different regions, that is they feature different dynamical properties, is not necessarily a sign of a different origin. Indeed, a massive merger, as the GSE one was, might have left debris on both radial and retrograde orbits, even within the solar neighbourhood \citep[][]{mori24}. On the other hand, the overlap of different substructures is not necessarily due to a common origin, as minor mergers of different accreted satellite galaxies might give rise to remnants that overlap with each other and with the one of the massive merger \citep[][]{jean-baptiste2017}. Since we cannot constrain the merger history of the Milky Way from the halo stars' dynamical properties only, we need additional information, that in this study is provided by the chemical abundances of these stars.

\section{Chemical properties}
\label{sec:abundances}

\begin{figure*}
    \centering
    \includegraphics[width=1.85\columnwidth]{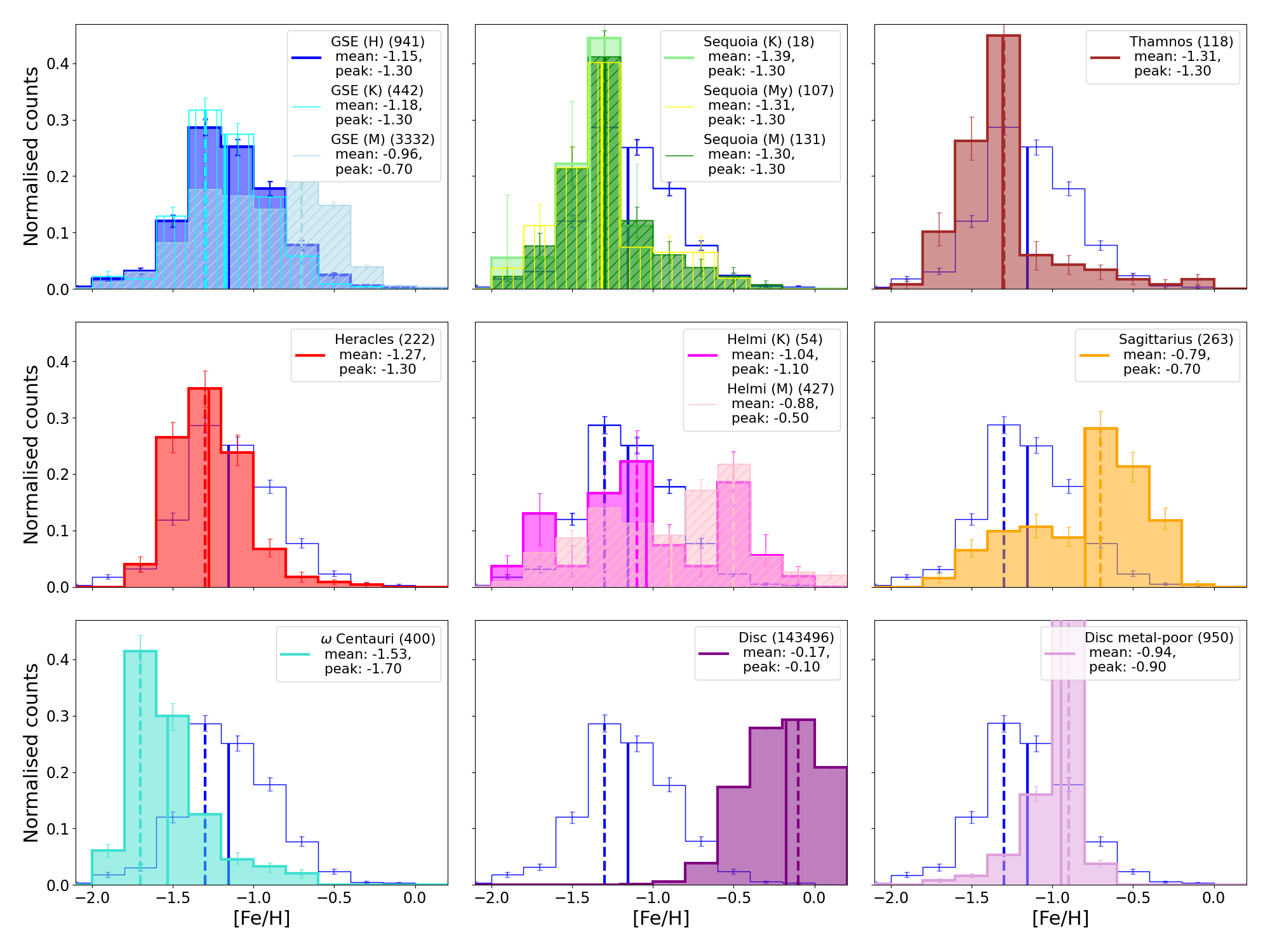}
    \includegraphics[width=1.9\columnwidth]{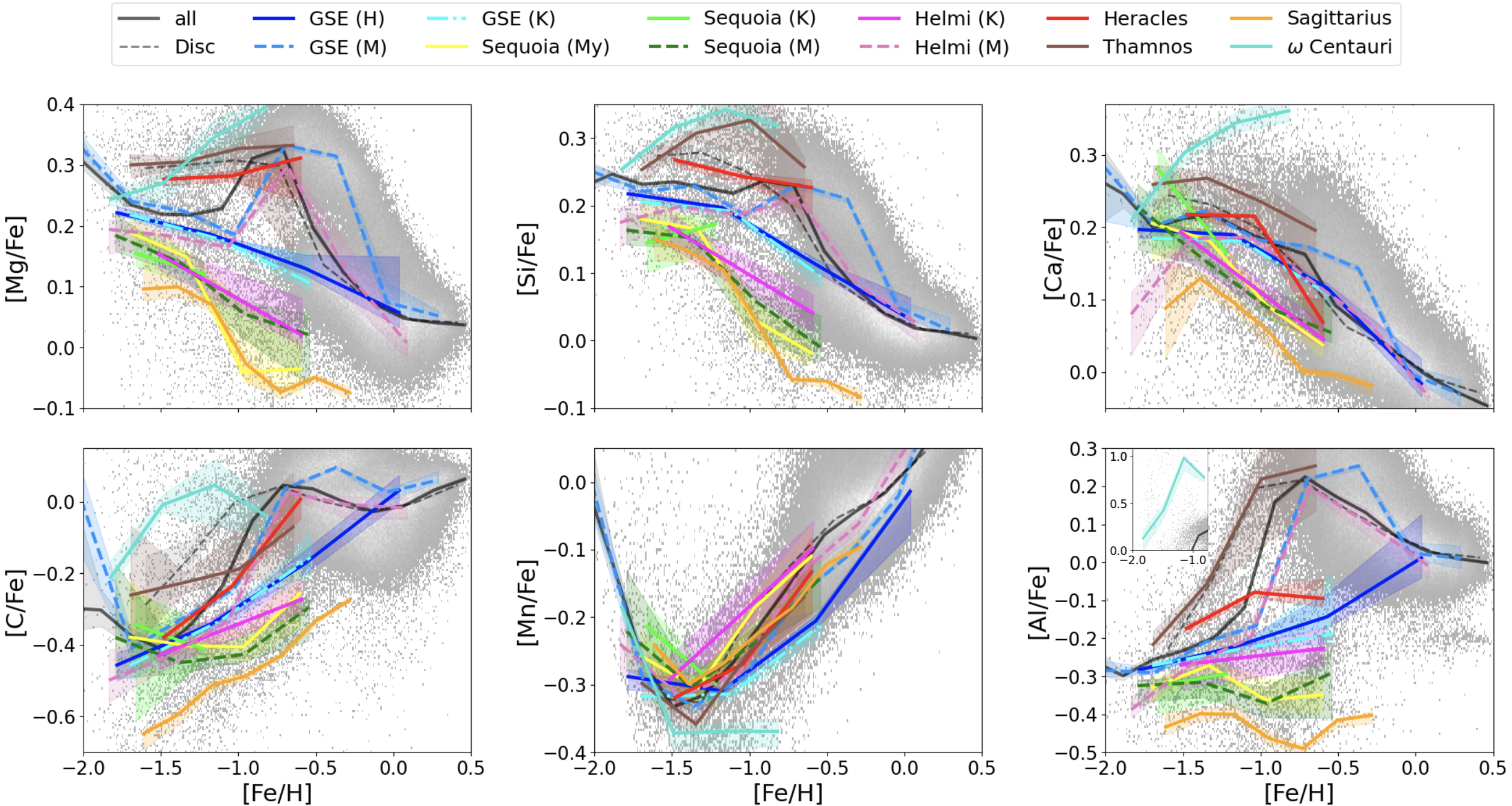}
    \caption{Upper panel: MDFs of the halo substructures in different colours with the same colour legend as in Figure \ref{fig:elz}. In the top left panel the GSE MDF (in dark blue/light blue/cyan) is shown for the three definitions considered (H23, K19 and M19). 
    In the following panels the blue GSE (H) MDF is compared to the ones of the other halo substructures, to the one of the disc, and finally to the metal-poor disc ([Fe/H] $<-0.8$) in the bottom right. The solid lines show the means, while the dashed ones the peaks (the values are reported in the legend). The error bars represent the confidence interval of 1000 bootstrap resamplings, where each resample draws stars with replacement and perturbs individual [Fe/H] values by their measurement uncertainties.
    Bottom panel: median distribution in the [Mg/Fe], [Si/Fe], [Ca/Fe], [C/Fe], [Mn/Fe] and [Al/Fe] vs. [Fe/H] spaces of the halo substructures, computed with a threshold of minimum three stars per bin. The sub-plot at the bottom right shows the median distribution of $\omega$ Cen with different [Al/Fe] and [Fe/H] ranges.
    The grey distribution shows the whole sample for reference. The shaded regions represent the bootstrap confidence interval on the running median, computed as above for each individual abundance ratio.
    }
    \label{fig:mdf}
\end{figure*}
We will begin chemically characterising the halo substructures, by showing their metallicity distribution functions (MDFs) in the upper panel of Figure \ref{fig:mdf}. The colour legend is the same as in Figure \ref{fig:elz} and in each panel we also report the GSE (H) MDF for comparison in blue (they are all normalised). The solid vertical lines show the mean of the distributions, whereas the dashed ones show their peak.
In the top left panel, we show the GSE MDF (in blue, cyan, and light blue for the H23, K19, and M19 selections, respectively):
its stars are on average metal-poor (<[Fe/H]> $= -1.15$). 
Through a comparison with the following panels, we can see that the MDFs of Sequoia (in green, yellow, and dark green for the K19, My19 and M19 selections, respectively), Heracles (in red) and Thamnos (in brown) all have similar peak values, being on average slightly more metal-poor \citep[e.g., see also][]{myeong2019, naidu2020}.
Then we show the MDFs of the Helmi Stream (in magenta for the K19 selection and in light pink for the M19 one), and the one of Sagittarius (in orange), which are on average more metal-rich and multiple-peaked. 
Moreover, we show the $\omega$ Cen MDF which is the most metal-poor (mean value: $-1.53$) and peaks at even lower values of metallicity. 
We finally show the MDF of the disc (in purple), and of the metal-poor disc sample (in pink), defined by the additional $\rm{[Fe/H]} \le -0.8$, which will be used in the following analyses.

Similar MDFs could hint at common origin, especially those that show very similar peak values along with similar shapes (e.g, GSE, Sequoia, Heracles, Thamnos), or hint at similar mass progenitors. However, different MDFs do not necessarily trace different accretion events, as a single massive satellite galaxy featuring a metallicity gradient can give rise to different MDFs in different regions of the $E-L_z$ space \citep[][]{mori24}. As a consequence, we are not able to discriminate between independent mergers yet.

We can then consider different abundance ratios, possibly tracing different chemical evolution histories. In the bottom panel of Figure \ref{fig:mdf}, we show the [Mg/Fe], [Si/Fe], [Ca/Fe], [C/Fe], [Al/Fe], and [Mn/Fe] abundance ratios as a function of [Fe/H]. The solid lines show the trend of the medians of the distributions of the various halo substructures, computed with a threshold of minimum three stars per bin. 
The colour legend is the same as in the previous figures. The grey 2D-histogram shows the distribution of the whole sample for reference. We can notice that the abundances of Sagittarius (in orange) are generally lower than in all the other substructures, with the exception of manganese. In the [Mn/Fe] vs. [Fe/H] the lowest median abundance is the one of $\omega$ Cen (in turquoise), which conversely is on average higher in all the other chemical spaces, in particular tracing a completely separate high-[Al/Fe] sequence. 
Among the other halo substructures, we can see that GSE (in blue/cyan/light blue), Sequoia (in green/dark green/yellow) and the Helmi Stream (in magenta/pink) usually trace similar sequences, occupying the regions of lower $\alpha$, C, Al and Mn compared to in situ components, typical of accreted populations and dwarf galaxies satellites of the Milky Way. Heracles and Thamnos feature instead on average higher abundances of $\alpha$ elements, carbon, and aluminum, generally connecting with the high-$\alpha$ thick disc sequence. We notice that the definitions of GSE, Sequoia and Helmi Stream from M19 -- as being more extended by definition -- are also closer to the high-$\alpha$ disc sequences with respect to the samples defined by K19 and H23, and in general reach higher values of metallicities. 

The analysis of the median trends of the halo substructures highlights some differences in their chemical characterisation (see the bottom panel of Figure \ref{fig:mdf}). However, computing the median of the abundances in iron bins, although allowing for a quick manifest comparison, on the other hand smooths out the specific patterns of the possible different populations comprised within the sample. 
In particular, given the broad redistribution in dynamical spaces of the massive merger debris and the consequential kinematic heating of the in situ component, the halo substructures defined by kinematic cuts are likely to feature a non-negligible amount of contamination from the GSE merger and the MW disc stars that have been heated to halo-like orbits. However, in the computation of the median, the high-$\alpha$ abundances characteristic of the in situ component might get lost, averaged with the low-$\alpha$ ones typical of GSE and, in general, of the accreted stars, possibly providing as the final result an intermediate $\alpha$ trend, depending on the percentage of contamination from GSE and the heated disc. Our aim is thus quantifying this amount of contamination, in order to verify whether these substructures are actually chemically distinct (on statistical level) from the GSE debris and from disc stellar populations, and in that case unravel their specific chemical patterns, ultimately constraining properties of their galaxy-progenitors and evolutionary history.

\section{Chemical analysis via Gaussian Mixture Models}
\label{sec:method}

\begin{figure*}
    \centering
    \includegraphics[width=2\columnwidth]{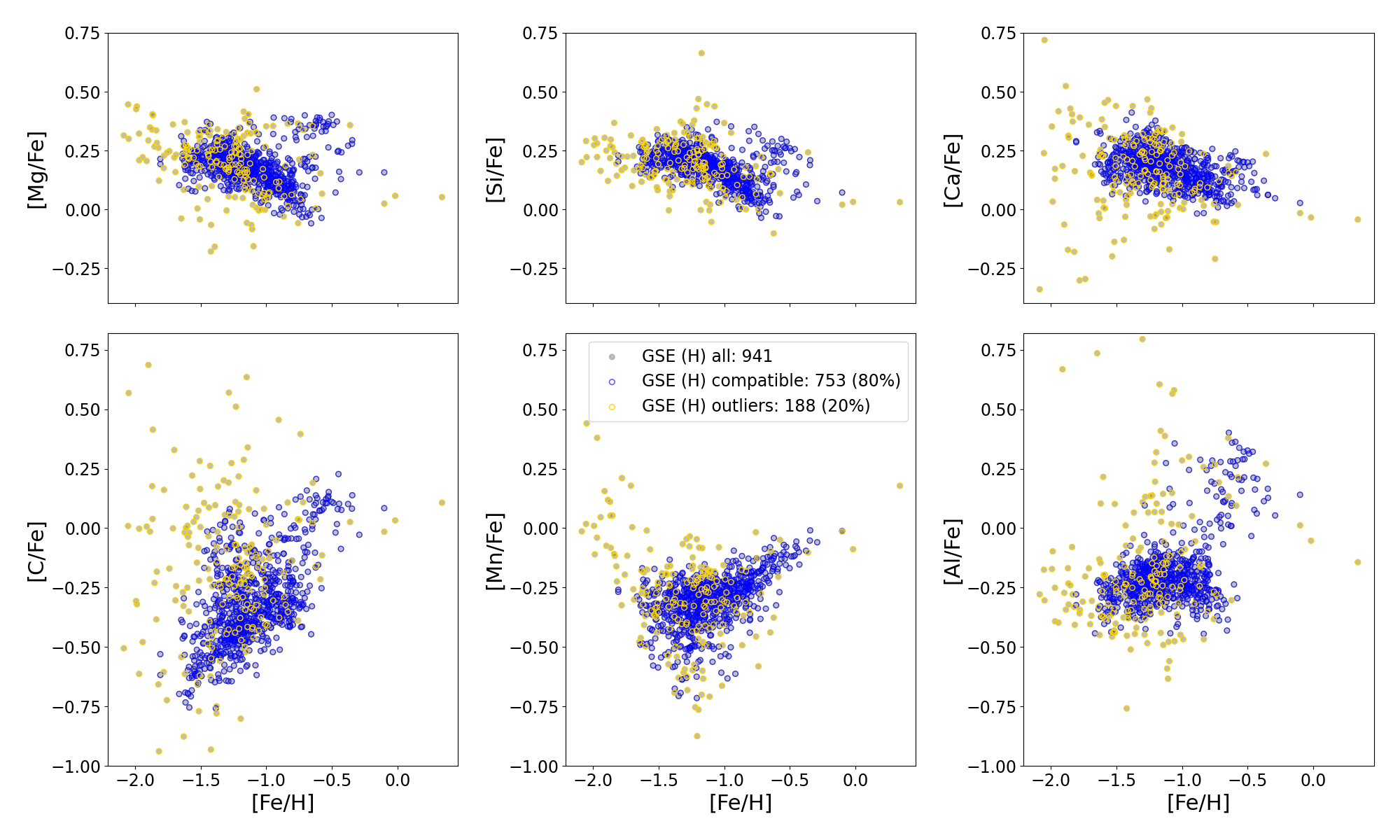}
    \caption{GSE distribution in the [Mg/Fe], [Si/Fe], [Ca/Fe], [C/Fe], [Al/Fe], and [Mn/Fe] vs. [Fe/H] spaces (in grey). The blue circles highlight the "pure GSE" reference sample used for the chemical comparison with the other halo substructures in the following.
    The yellow circles highlight instead the outliers of the probability distribution of stars of belonging to the GSE GMM of 4 components, that is the lowest 20th percentile in this case.}
    \label{fig:gmmGSE}
\end{figure*}

The unravelling of the (possible) intrinsic chemical patterns of the halo substructures requires to quantify (and clean out) the contamination from the last significant merger of the MW and from the kinematically heated in situ disc stars, described in the previous Section \ref{sec:abundances}. As the halo substructures are defined by pure kinematic cuts (with the exception of Heracles), we exploit their abundances to investigate their connection with the other halo components. We tackle the chemical comparison of the halo substructures with respect to GSE and the heated disc by means of Gaussian Mixture Models (GMM). 

GMM is a probabilistic modelling technique that assumes that the distribution of the dataset is a combination of a finite number of Gaussian components. GMM is particularly well-suited for identifying substructures in multidimensional data, where single-Gaussian assumptions may not accurately capture the complexity of the distribution. The number of Gaussians and their parameters are not known a priori; the optimal model is selected on the basis of the Bayesian Information Criterion (BIC). Lower BIC values indicate a better model while penalising excessive components. This method balance model complexity (number of Gaussians) with goodness-of-fit to prevent overfitting. 
Compared to alternative clustering techniques such as k-means, which assigns each data point to one cluster, GMM provides a soft clustering approach, assigning probabilities of belonging to different clusters, thus allowing for overlapping distributions and incorporating uncertainties.

The choice of using GMMs is driven by the fact that to perform the analysis we first of all want to take into account several different chemical abundances at the same time, to probe different nucleosynthetic channels, and discriminate between different star formation and chemical evolution histories. We considered the following abundances: [Fe/H], [Mg/Fe], [Si/Fe], [Ca/Fe], [C/Fe], [Al/Fe] and [Mn/Fe], as motivated in Section \ref{sec:data}.
In the second place, we also want to compare the whole distributions of a given element with respect to iron  rather than average values, because - knowing that these kinematic-defined samples imply a certain level of contamination - we do not want to lose the information of the different populations comprised that may be traced by the elements' patterns but not by the average. The use of GMMs allows us in this way to encompass the global chemical evolution of the halo substructures, imprinted in the distributions of the abundances, at the same time enabling us to account for all the chosen elements at once, not having to project in 2-dimensional spaces for the analysis. 

Our aim is to evaluate the chemical compatibility of the various halo substructures with the main accreted component (GSE) and with the main in situ component (the disc).
In order to do so, we considered the GSE distribution in the 7-dimensional chemical space as our training sample, and we compared all the other substructures' samples with it. We performed the analysis with two definitions of GSE, that is (H), (K), and (M), as to verify whether the results were depending on the kinematic selection assumed, and we tested different thresholds for the GMM. 

In Figure \ref{fig:gmmGSE}, we show the GSE (H) distribution in the [Mg/Fe], [Si/Fe], [Ca/Fe], [C/Fe], [Al/Fe], and [Mn/Fe] vs. [Fe/H] spaces. 
The optimal number of Gaussian components for GSE given by the minimisation of the BIC is four. The blue circles in the figure highlight the bulk of the distribution (combining the four components), defining our GSE reference sample ("pure GSE"). 
The yellow circles are the outliers of the GSE distribution, that is the stars with the lowest probability density of belonging to the four components of the GSE GMM (in Fig. \ref{fig:gmmGSE}, we set the threshold to the 20th percentile).
We can see that most of the outliers are low-metallicity stars ($\rm[Fe/H] \le -1$), that are very scattered and feature in general higher abundances of carbon, manganese, and aluminium over iron (bottom row), while being still consistent with the bulk of the pure GSE in the $\alpha$ elements (upper row).

We then compared the other halo substructure samples with the pure GSE reference one, evaluating the probability density that any of their stars belonged to the GMM of GSE. The ones with a probability higher than the assumed threshold were considered GSE-compatible stars; conversely, they were considered not compatible.
In this way, we obtain a percentage of chemical compatibility of the various halo substructures with respect to GSE. We associated an uncertainty with these fractions by bootstrap resampling the data a thousand times, considering the uncertainties of the chemical abundances given by APOGEE with a Monte Carlo approach. We finally computed the means and standard deviations of the realisations.

The same approach has been used to quantify the chemical compatibility of the different halo substructures with the kinematically heated disc. 
Because the chemical abundances of the heated disc are expected to be included in the ranges spanned by the disc \citep[see, e.g.][]{dimatteo2019, belokurov20}, to quantify this compatibility, we made use of the chemical abundances of the disc itself, as defined in Sect. \ref{sec:abundances}.

\section{Chemical compatibility of the halo substructures}
\label{sec:results}

In this Section, we present the results of the chemical comparison analysis of the various halo substructures with respect to GSE (Sec. \ref{sec:res:gse}) and the heated disc (Sec. \ref{sec:res:disc}), performed by means of GMMs considering the following elements: [Fe/H], [Mg/Fe], [Si/Fe], [Ca/Fe], [C/Fe], [Al/Fe], and [Mn/Fe]. We finally show in Sec. \ref{sec:res:rest} the pattern of the cleaned halo substructures, that is the stars that are not chemically compatible neither with GSE nor with the heated disc, and that thus could hint new meaningful details about the evolution of the Milky Way.

\subsection{Comparison with the accreted component: GSE}
\label{sec:res:gse}

The fractions of chemically compatible stars within each halo substructure with respect to GSE are reported in the different columns of Table \ref{tab:comp}, along associated uncertainty, computed by bootstrap resampling the data ($n_{boot}= 1000$) as described in Section \ref{sec:method}. 
Furthermore, we list the results obtained considering different threshold values for the accepted probability density, which are 5, 10, 15 and 20 percentiles. The fractions vary accordingly, but the ranking of the most chemically-compatible substructures is not affected.
Finally, we show the chemical comparison with respect to the three different definitions of GSE considered in this work (H, K, and M), the results of which are generally consistent with each other within the uncertainties. 

The extensive abundance distributions of stars of each halo substructure chemically compatible with pure GSE and not are reported in the upper panels of the figures in the Appendix \ref{appendix:comp}. 

In Figure \ref{fig:comp_20_1000} we show for each of the halo substructures considered in this study (reported on the x-axis) the relative percentage (and uncertainty) of chemical compatibility with respect to GSE, in blue, cyan, and light blue when considering the GSE (H), (K), and (M) definitions. These percentages were obtained when a threshold of 20\% was considered, so in this case the upper limit for the compatibility is 80\%. For this reason, we report on the right y-axis the percentages normalised to 80\%, to highlight the absolute chemical compatibility. The choice to assume a threshold of 20\% to quantify the possible contamination by GSE for the various substructures was driven by the objective of being as conservative as possible, of selecting the very bulk of the GSE's distribution (the "pure GSE" sample; see Sec. \ref{sec:method}) and of finding results independent from its outskirts.

\begin{figure}
    \centering
    \includegraphics[width=1\columnwidth]{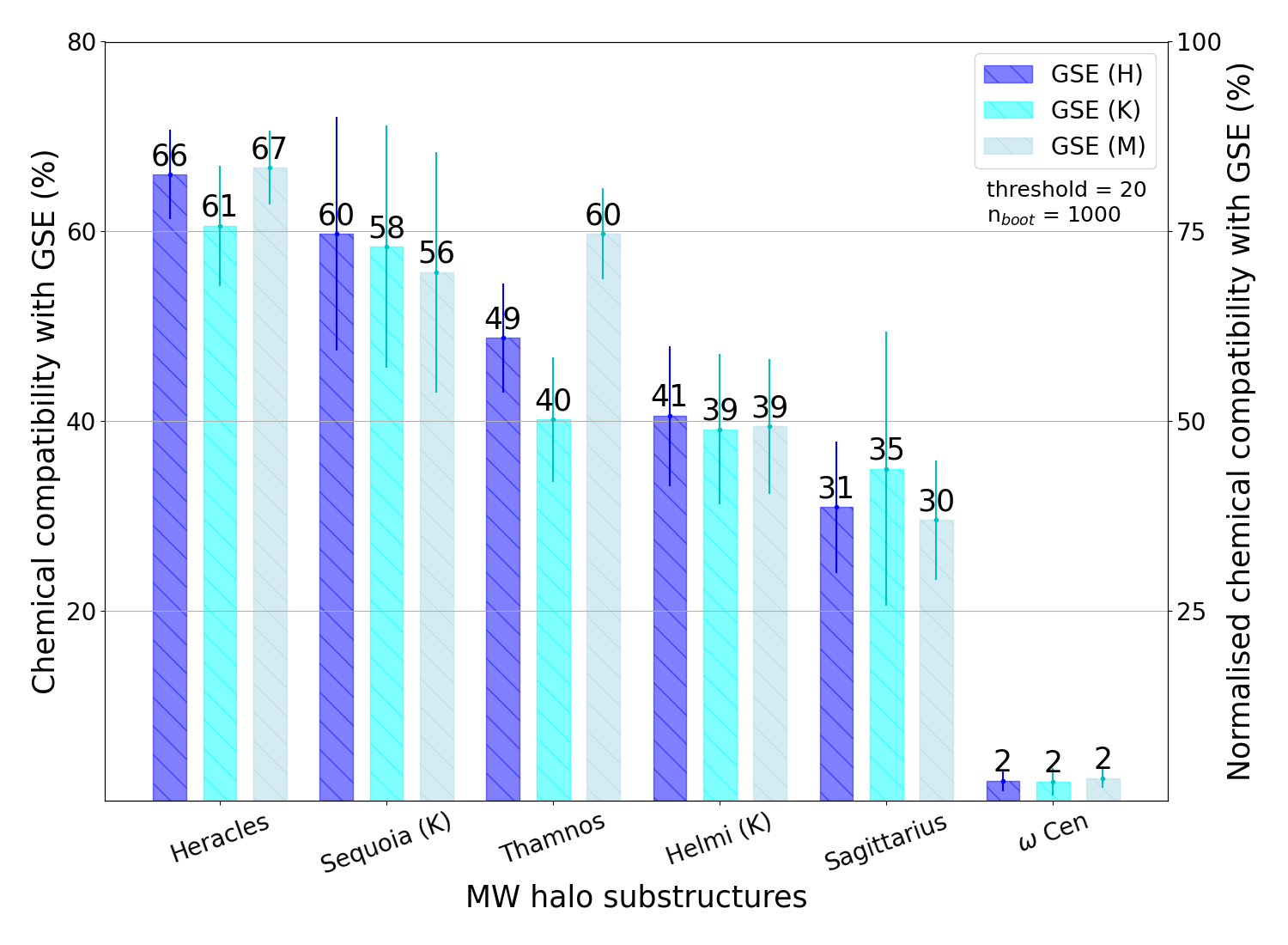}
    \caption{Fractions (\%) of chemically compatible stars of the Milky Way halo substructures with respect to GSE (H) in blue, GSE (K) in cyan, and GSE (M) in light blue. The relative uncertainties are obtained by bootstrapping the data a thousand times, accounting for the observational uncertainties. 
    The fractions are computed assuming a threshold equal to the 20th percentile of the probability density, thus the chemical compatibility normalised to 80 percent is reported on the right y-axis.} 
    \label{fig:comp_20_1000}
\end{figure}

From the fractions of chemically compatible stars, we can draw the following outcomes:
\begin{itemize}
    \item  Heracles and Sequoia (K) feature the highest fraction of chemically compatible stars with GSE (H), 66\% and 60\% respectively (out of the 80\% maximum possible), with comparable values when considering the GSE (K) and (M) definitions. In particular, for Sequoia, stars compatible with GSE redistribute quite uniformly over the whole extent of GSE abundances (see the upper panel of Fig. \ref{fig:sequoia_cut}). These two facts suggest that the bulk of its stars (a fraction of 75\% in absolute terms) might have had a common origin with GSE. It is unlikely that completely independent systems had such similar chemical patterns, a fortiori given that they are supposed to have been accreted at different redshifts and with different mass ratios \citep[][]{kruijssen20}. In any case, it is very hard with current data to distinguish between the nowadays chemical distributions of Sequoia from that of GSE. 
    With higher resolution chemical abundances, \citet{matsuno22a} and \citet{ceccarelli24} found a lower contamination from GSE within their Sequoia samples.
    The Heracles distribution, instead, although featuring a higher chemical compatibility with GSE with respect to Sequoia, does not cover completely that of GSE, being generally shifted at higher $\alpha$ abundances (see the upper panel of Fig. \ref{fig:heracles_cut}). However, we remind that the Heracles definition is based on abundance criteria as well, artificially removing Al-rich and low Mg/Mn stars.
    \item Thamnos showcases a high percentage of compatibility as well ($\sim$50\%), meaning that one every two stars in Thamnos is chemically compatible with the bulk of GSE (H) (compatibility higher than 60\% with the pure GSE). However, in this case, there is a higher discrepancy when considering different GSE definitions, with an even higher compatibility fraction when considering the GSE (M) definition ($\sim$60\%, 75\% with pure GSE)). This discrepancy will be clearer when putting it in context with the comparison with the heated disc; yet again in both cases the scenario of a shared formation with GSE is  plausible, this time possibly comprising at least another population with a different origin.
    \item The Helmi Stream exhibits a lower percentage of GSE (H) compatible stars, but still equal to 2 every 5 ($\sim$40\%, 50\% with pure GSE), consistently with the GSE (K) and (M) values. In this case, even though there might be a substantial presence of GSE stars within this sample, we expect a significant contribution from at least another stellar population as well. 
    \item Sagittarius presents approximately one third of stars chemically compatible with GSE, consistently in all the definitions. This is interesting because in this case we know already that these must be different systems as we can still observe the elongated spheroid remnant of the dwarf spheroidal galaxy of Sgr and its tidal stream with leading and trailing branches for tens of kpc along their quasi-polar orbit. As a consequence, the non-negligible fraction of chemical compatibility that we find between Sgr and GSE could actually hint that GSE may comprise stars coming from the Sagittarius Stream itself. This would be also consistent with the fact that we can find Sgr stripped stars at comparable distances from the Sun and that the galaxy is predicted to have passed within the solar neighbourhood contributing to the formation of the stellar halo from both test-particle orbits and N-body simulations \citep[][]{lawmajewski09, laporte19}. On top of that, if we take a look at the $E-L_z$ space (Fig. \ref{fig:elz}), we can also notice that the distribution of Sgr partly overlaps with that of GSE in the high energy region ($1.4 \le E \le 1.1$), corroborating the contamination of Sgr stars within the kinematically defined GSE sample.
    \item $\omega$ Cen is the sample that shows the lowest compatibility with GSE in our study (2\%). This is a surprising result, because it is suggested to be the nuclear star cluster of GSE (see \citealp{massari2019, callingham22, limberg22}). In fact, the two nuclear star clusters for which we have detailed chemical abundances, namely M54 and the Galactic nuclear star cluster, show a remarkably similarity with their host galaxies Sagittarius \citep[see][]{pagnini25} and the Galactic bulge/inner disc itself \citep[][]{ryde25, nandakumar25}, respectively. 
\end{itemize}

\subsection{Comparison with the in situ component: the disc}
\label{sec:res:disc}

\begin{figure}
    \centering
    \includegraphics[width=1\columnwidth]{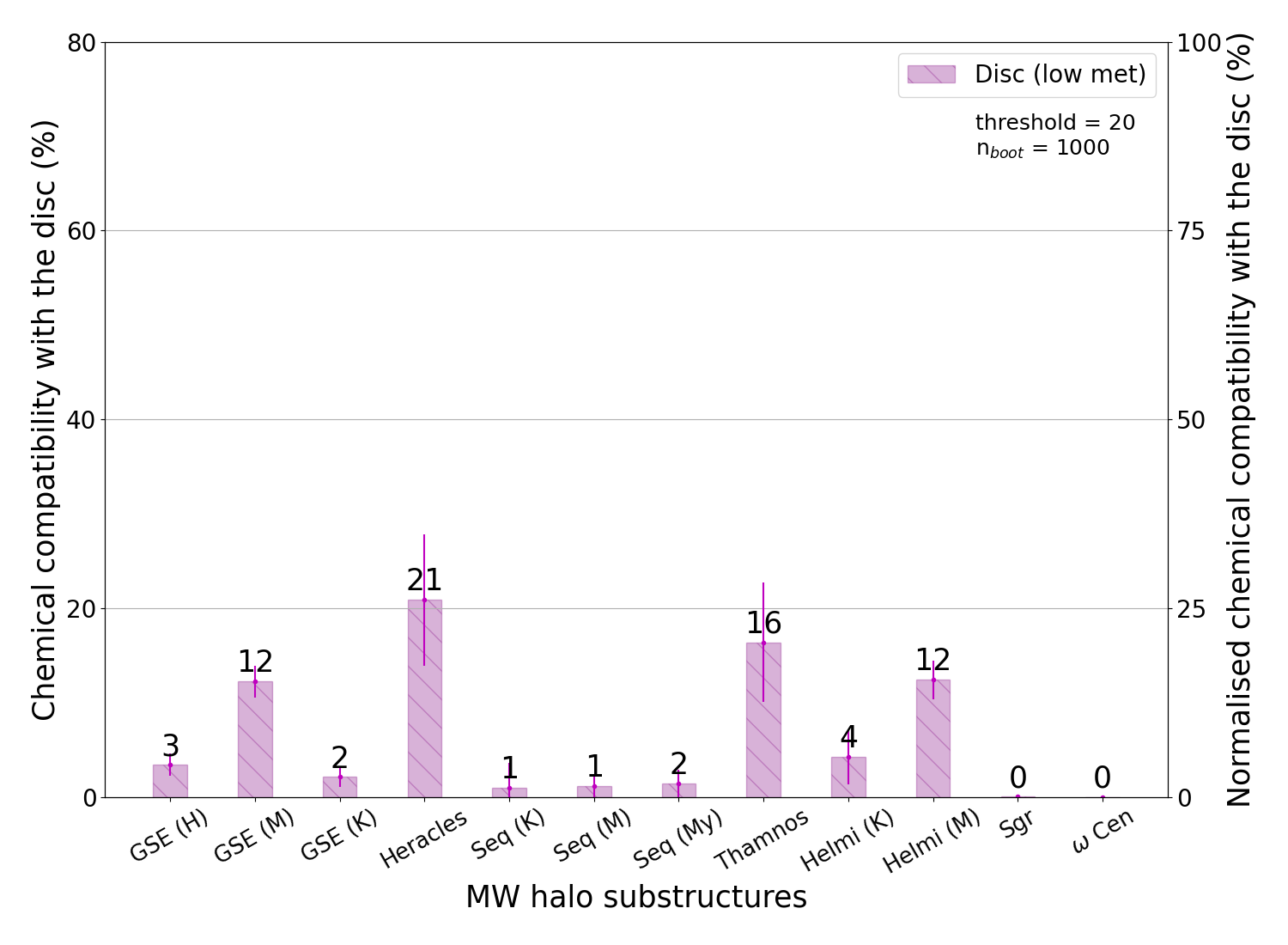}
    \caption{Fractions (\%) of chemically compatible stars of the Milky Way halo substructures with respect to the metal-poor disc ($\rm{[Fe/H]} \le -0.8$). The relative uncertainties are obtained by bootstrapping the data a thousand times, accounting for the observational uncertainties.
    The fractions are computed assuming a threshold equal to the 20th percentile of the probability density, thus the chemical compatibility normalised to 80 percent is reported on the right y-axis.} 
    \label{fig:compatibility_disc}
\end{figure}

In a complementary way with respect to the comparison with the main accreted component (i.e., GSE), we want to chemically compare the halo substructures to the main in situ component, that is the disc, to quantify the fraction of these substructures which may be contaminated by heated disc stars -- that is stars originally born in the disc of the MW, and subsequently heated by one or several merger events to halo kinematics \citep[e.g., ][]{dimatteo2019, belokurov20}. Because this heated disc is expected to be made of stars with the chemical abundances also found in present-day disc stars, to estimate its contribution to the halo substructures, we quantify the chemical compatibility with the disc itself.

As Figure \ref{fig:mdf} shows, many substructures have chemical tracks which tend to converge to those of the disc at low metallicities. To ensure that we had sufficient statistics to construct a robust GMM model for low-metallicity disc stars, we therefore defined a disc sample, selecting stars with metallicity $\rm{[Fe/H]} \le -0.8$ and circularity $\eta \ge 0.8$. 

For each halo substructure, in the different columns of Table~\ref{tab:comp_disc} we report the fractions of stars chemically compatible with the metal-poor disc along with the total number of stars and their associated uncertainty ($n_{boot}= 1000$). 
As for the case of the comparison with GSE (Table~\ref{tab:comp}), we list the results obtained considering different threshold values for the accepted probability density, that are 5, 10, 15 and 20 percentiles. The fractions vary accordingly, but the following results hold true. 

In Figure \ref{fig:compatibility_disc} we summarise the relative percentages (and uncertainties) of chemically compatible stars with respect to the metal-poor disc for each of the halo substructures considered in this study (reported on the x-axis).
These percentages were also obtained assuming a threshold of 20\%, so the upper limit for the compatibility is 80\%. For this reason, we report on the right y-axis the percentages normalised to 80\%, to highlight the absolute chemical compatibility. As for the corresponding abundance distributions of chemically compatible -- and not -- stars of each halo substructure with respect to the metal-poor disc, we refer the reader to Appendix \ref{appendix:comp} (bottom panels of the figures).

From the chemical comparison with the metal-poor disc, we might divide the MW halo substructures into two groups: 
\begin{itemize}
    \item one made up of substructures that have little in common with the disc (chemical compatibility $\le$ 5-10\%). The halo debris with the lowest chemical compatibility with the metal-poor disc are: {Sequoia (1\% both in the H and K definitions, 2\% in the M one), the Sagittarius Stream (0\%), and $\omega$ Cen (0\%)}. Therefore, these samples can be fairly considered clean from the presence of metal-poor heated disc stars. Then we also have samples with a low but not null disc contamination: GSE (H) and (K) (5\% and 4\%, respectively) and the Helmi Stream (K) (4\%). 
    \item the other one comprises the systems that feature instead considerable chemical compatibility with the disc in the low-metallicity regime (of the order of 15-25\%). This group includes Heracles (21\%) and Thamnos (17\%), but also the M19 definitions of GSE (15\%) and Helmi Stream (12\%). We notice a correlation between the high chemical compatibility with the metal-poor disc and the high $\alpha$-elements' abundances: the "high-$\alpha$" halo substructures are generally found to be similar also in the other chemical abundances to the low metallicity tail of the disc by the GMM analysis.     
\end{itemize}

\subsection{The cleaned substructures} 
\label{sec:res:rest}

Once contamination from the massive merger of GSE and from the heated disc has been identified and quantified in each halo substructure, the contaminant stars can be removed from each substructure to highlight possible intrinsic chemical patterns in the remaining cleaned substructures. The conclusions drawn from their analysis are the following. 

\begin{figure*}
   \centering
   \includegraphics[width=2\columnwidth]{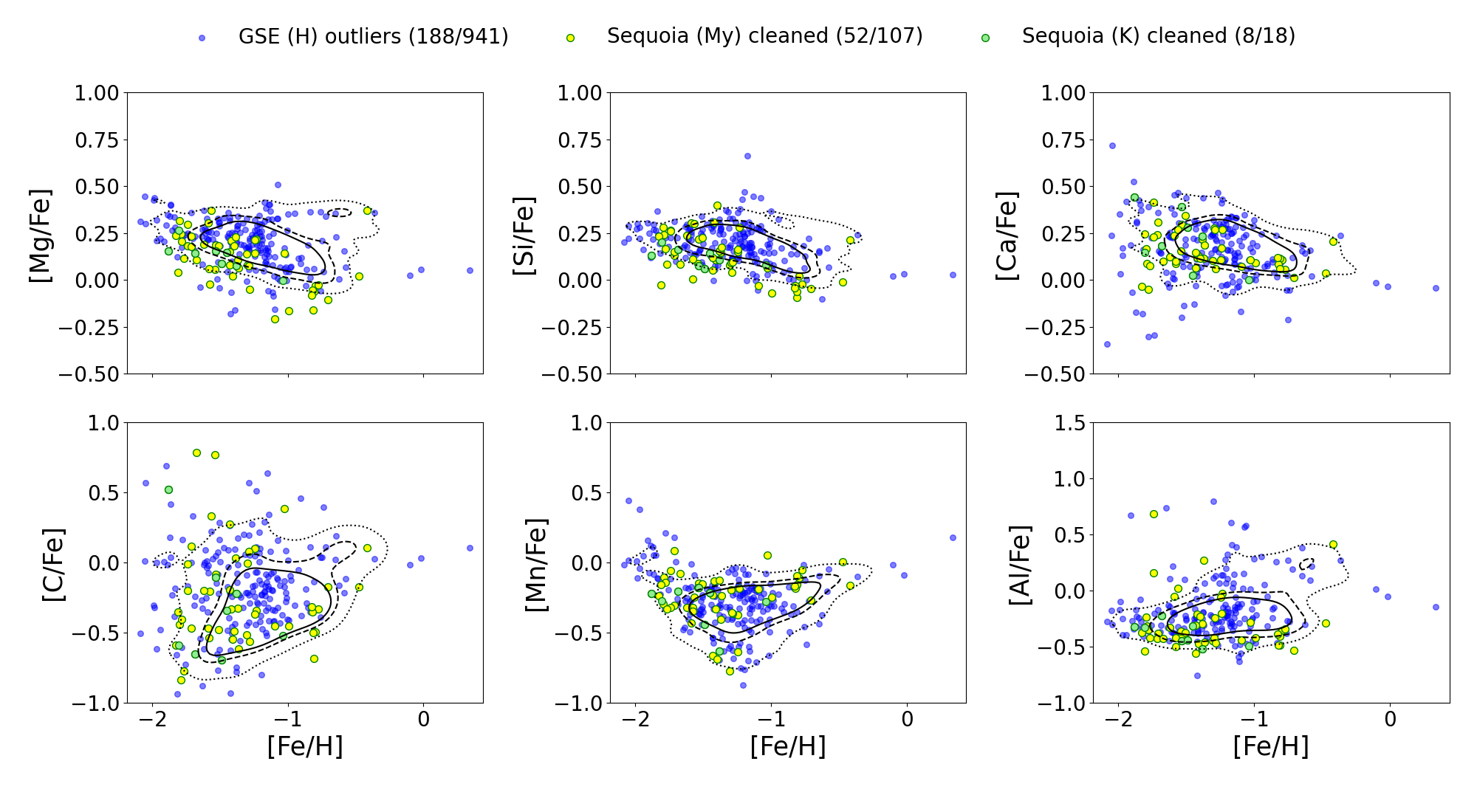}
   \caption{Abundance distribution in the [Mg/Fe], [Si/Fe], [Ca/Fe], [C/Fe], [Al/Fe], and [Mn/Fe] vs. [Fe/H] spaces for the cleaned Sequoia (K) (in green) and Sequoia (My) (in yellow) samples, compared to the outliers of GSE (H) (in blue).
   The black contours show the 68\%, 80\% and 95\% of the pure GSE distribution for reference.}
   \label{fig:outliers}
\end{figure*}

\begin{figure*}
   \centering
   \includegraphics[width=2\columnwidth]{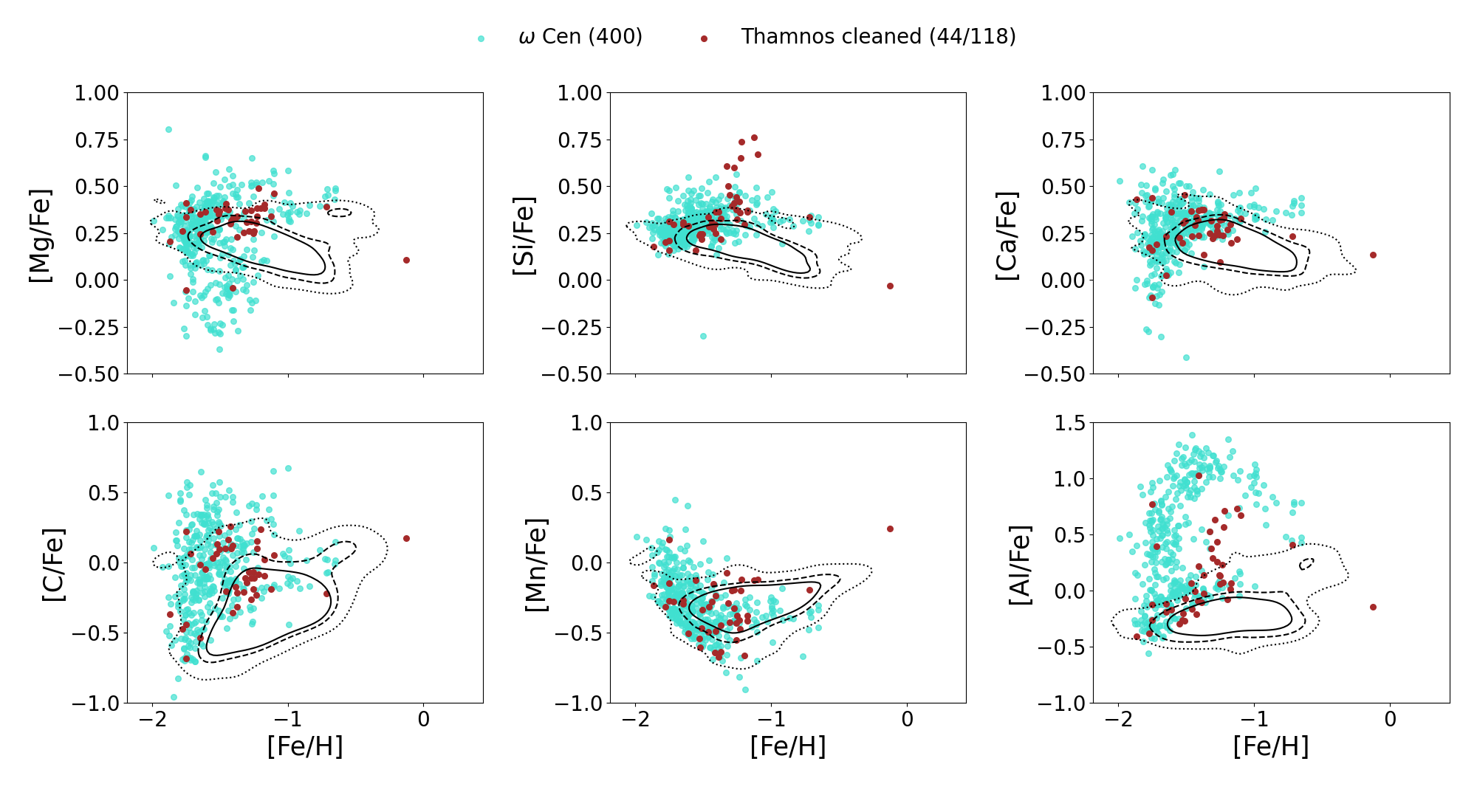}
   \caption{Abundance distribution in the [Mg/Fe], [Si/Fe], [Ca/Fe], [C/Fe], [Al/Fe], and [Mn/Fe] vs. [Fe/H] spaces for the cleaned Thamnos (in brown) and $\omega$ Cen (in turquoise). The black contours show the 68\%, 80\% and 95\% of the pure GSE distribution for reference.}
   \label{fig:thmanos-ocen}
\end{figure*}

\begin{figure*}
   \centering
   \includegraphics[width=2\columnwidth]{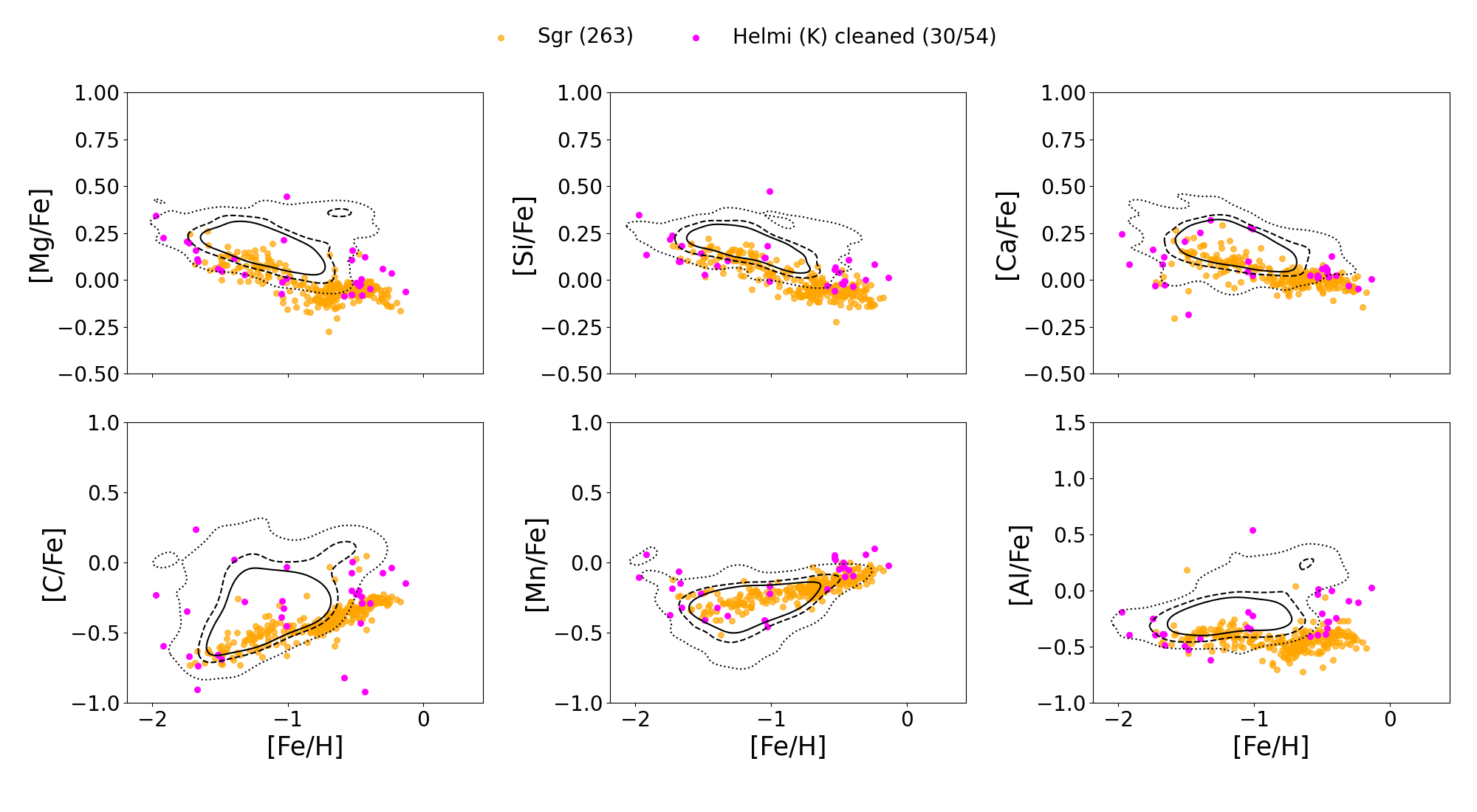}
   \caption{Abundance distribution in the [Mg/Fe], [Si/Fe], [Ca/Fe], [C/Fe], [Al/Fe], and [Mn/Fe] vs. [Fe/H] spaces for the cleaned Helmi Stream (in magenta) and Sagittarius Stream (in orange). The black contours show the 68\%, 80\% and 95\% of the pure GSE distribution for reference.}
   \label{fig:helmi-sgr}
\end{figure*}

\begin{itemize}
    \item In the Heracles sample, the abundances of the vast majority (at least 80\%\footnote{This percentage has been derived by adding the percentage of Heracles stars which are compatible either with GSE (H) or with the metal-poor disc, and subtracting to this sum the disc contamination of GSE (H), which is of the order of 5\% (see Tab.~\ref{tab:comp_disc} and Fig.~\ref{fig:compatibility_disc}).}) of the stars can be explained as a combination of two different stellar populations: the GSE debris and metal-poor disc stars. Thus, very little room is left for interpretation of the substructure cleaned from these two components.

    \item As long as Sequoia is concerned, instead, beyond the high chemical compatibility with GSE, we still have a small but significant group of stars (about 39\% of the sample) that result as being compatible neither with GSE nor with the disc. In Figure \ref{fig:outliers} we show the distribution in the chemical spaces for the cleaned Sequoia (K) and Sequoia (My) with respect to GSE (H), with the same colour code as in the previous figures. For reference, we show the 68\%, 80\%, and 95\%  contours of the pure GSE probability density distribution (in black).
    If we take a closer look at the cleaned Sequoia (in green/yellow) and we visually compare them to the GSE's outliers (in blue), we can actually notice that many of the green/yellow points are still distributed within the regions of the chemical spaces where also the blue points lie. This suggests that the threshold assumed for the modelling cut out the same kind of populations from both samples, meaning that a significant part of the cleaned Sequoia could actually be compatible with the outliers of GSE. As a consequence, we argue that the chemical compatibility between the Sequoia and the GSE samples is at least as high as 60\%, strongly hinting at a common origin for GSE and Sequoia. 
    
    In this interpretation, Sequoia would be made of metal-poor\footnote{As shown in Fig.~\ref{fig:mdf} and by others \citep[e.g., ][]{myeong18}, Sequoia has a mean metallicity which is lower than that of GSE.} stars once part of the GSE progenitor, and which were subsequently lost on retrograde orbits,  during the early phases of its accretion into the Galaxy. Note that this interpretation is valid if the GSE progenitor had initially a metallicity gradient, as predicted by simulations \citep[e.g.,][]{koppelman20, amarante22, khoperskov23b, khoperskov23d, mori24}.
    
    We emphasise, however, that among the stars in Sequoia (My) which have low probability to be chemically compatible with GSE, some have chemical abundances systematically different from that of the GSE outliers. This is mainly visible in Fig.~\ref{fig:outliers} at [Fe/H] $\gtrsim -1$, where some (5-10) stars (in yellow) have [Mg/Fe] and [Si/Fe] lower than those of GSE outliers in the same [Fe/H] interval. This could be explained again by abundance gradients within GSE, due to the fact that star formation and chemical enrichment are more efficient in the centres of galaxies compared to their outskirts. Consistently with coming from GSE outskirts, Sequoia stars are not only on average on higher energies and more retrograde orbits (see Fig. \ref{fig:elz}) and are more metal-poor, being the first stars to be lost, but also its cleaned sample features on average lower [Mg/Fe] and [Al/Fe] abundances as expected for stars coming from regions of less efficient chemical enrichment \citep[see upper and middle panels in Figure 3 of ][]{skuladottir25}. 
    Given the limited statistics available, it is difficult to conclude definitely on this point, other than to point out that there is the possibility of a 10\% at most of stars in the Sequoia sample have an origin other than GSE. 

    \item Regarding Thamnos, despite being significantly compatible with both GSE and the disc, its cleaned sample features some peculiar trends in its abundances, different from GSE/disc and the other substructures. In particular, we can observe a prominent branch in high [Al/Fe] that is unusual in common satellite dwarf galaxies of the MW, but rather typical of globular clusters. We then decided to compare the cleaned Thamnos with $\omega$ Cen, as they are also very close (partly overlapping) in the $E-L_z$ space (see Fig. \ref{fig:elz}).
    In Figure \ref{fig:thmanos-ocen}, we show the distribution in the chemical spaces of the cleaned Thamnos (in brown) and of $\omega$ Cen (in turquoise). 
    We can indeed see that the majority of the stars of the cleaned Thamnos lie within the $\omega$ Cen chemical distribution in all the abundance spaces. 
    However, on average, the cleaned Thamnos is shifted at higher metallicities than the bulk of $\omega$ Cen stars and features smaller spreads. 
    If there is indeed a link between these two systems, that is that the cleaned Thamnos might be associated with part of the stars lost by the galaxy accreted with $\omega$ Cen, it suggests that the progenitor of $\omega$ Cen must have also originally had a metallicity gradient. 
    Furthermore, we can observe a metal-rich star of Thamnos ([Fe/H] $\sim -0.1$) isolated from both distribution, but we recall that the comparison with the disc is done only for [Fe/H] $\le -0.8$, so that the cleaned Thamnos' metal-rich star could likely be a heated disc star (see bottom panel of figure \ref{fig:thamnos_cut}).

    \item Finally, as for the Helmi Stream, we already mentioned its chemical compatibility with GSE and the disc: both are likely present in the Helmi Stream, but they do not make up the whole sample. We then considered its cleaned sample and compared it this time with the Sagittarius dSph and stream chemical distribution. This choice was driven by the fact that they are very close in the $E-L_z$ space, but also in the $L_{\perp}-L_z$ space, by which the Helmi Stream is defined (see Figure \ref{fig:elz} and bottom panels of Figure \ref{fig:helmi-sgr-appendix} in Appendix \ref{appendix:helmisgr}), partly overlapping in both spaces, especially in the high-$E$ and high-$L_{\perp}$ regions. 
    In Figure \ref{fig:helmi-sgr}, we show the distribution in the chemical spaces of the cleaned Helmi Stream compared to the Sagittarius distribution. We can see that the cleaned Helmi Stream (in magenta) overlap quite nicely with the Sgr distribution among the chemical spaces.  
    Moreover, we explored the spatial distribution of the cleaned Helmi Stream with respect to that of the Sagittarius (see $lon-lat$, $x-y$, $x-z$, and $y-z$ spaces in the upper two rows of Figure \ref{fig:helmi-sgr-appendix} in Appendix \ref{appendix:helmisgr}). Especially in the $y-z$ and space, it is possible to appreciate how the cleaned Helmi Stream is actually sitting in the middle of the Sagittarius Stream.
    The combination of these spatial, dynamical and chemical arguments hints in the direction of a significant contamination from Sagittarius within the Helmi Stream.
    Finally, we point out that the three cleaned Helmi Stream's stars that are the most metal-rich and high-$\alpha$ are probably disc stars, as they overlap in all chemical spaces with the in situ distribution, but the GMM-based comparison with the disc had been done only for metallicity [Fe/H] $\le -0.8$.
    These are stars would have likely been recognised by the GMM model as disc contaminants if we had used a higher threshold in metallicity for building the GMM model of the disc than what we have chosen.
    
\end{itemize}

\section{Discussion}
\label{sec:discussion}

\begin{figure*}
   \centering
   \includegraphics[width=1\columnwidth]{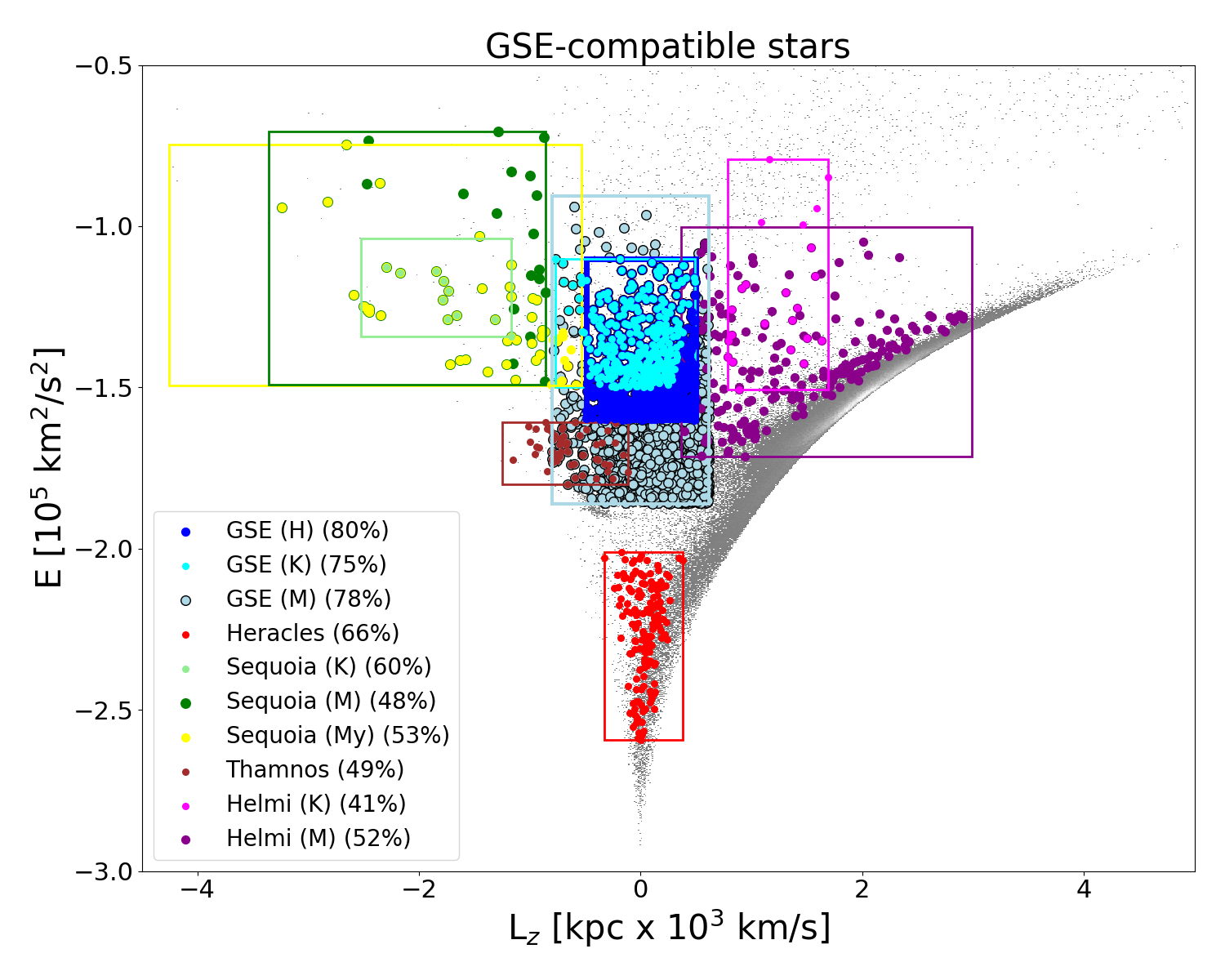}
   \includegraphics[width=1\columnwidth]{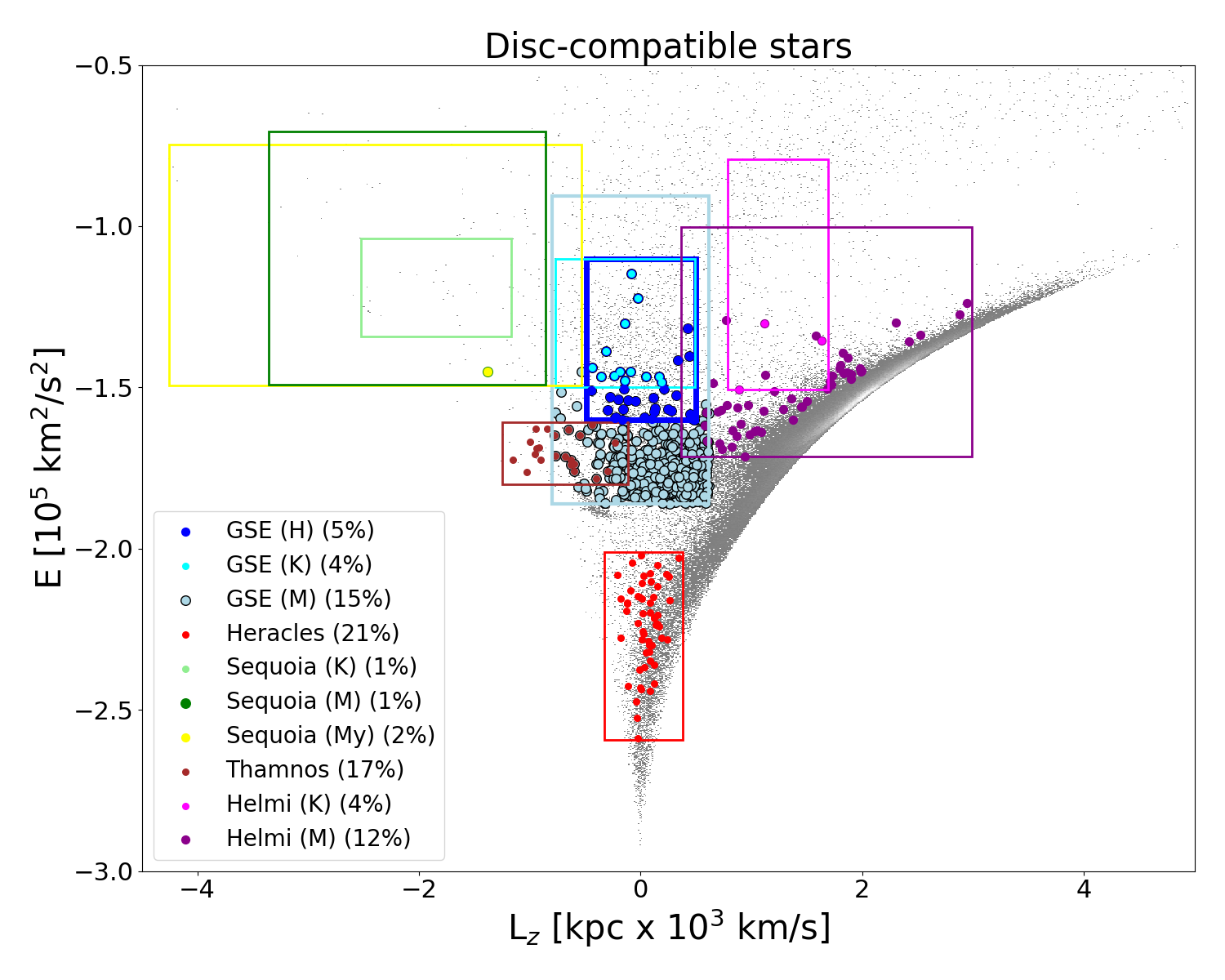}
   \caption{Distribution in the $E-L_z$ space of all the stars that are chemically compatible to GSE (H) (left panel) and the metal-poor disc (right panel), within each halo substructure (the number of the compatible stars are reported in the legend). 
    For reference we show the kinematic ranges spanned by these substructures by definition as rectangles with the same colour legend as the points, along with the GSE (H) and (M) definitions in blue and cyan respectively.}
   \label{fig:compatibility_elz}
\end{figure*}

In the previous Section, we have chemically compared the kinematic substructures in the Milky Way stellar halo with the GSE and the metal-poor disc with GMMs, in order to quantify the contamination from these two components within the system's samples. We have been able to identify the stars within each substructure that might be related to GSE and/or to the disc, and once we have cleaned them out from the substructures' samples we indeed observed peculiar hidden chemical patterns that might have outshone different formation scenarios for the systems analysed. What about the GSE/disc-compatible stars then? What are their dynamical properties?

To shed light on the potential dynamical distribution of GSE within the $E-L_z$ space, we show all GSE chemically compatible stars of the halo substructures that we discovered from the analyses with GMMs in the left panel of Figure \ref{fig:compatibility_elz}. For reference, in blue/cyan/light blue rectangles we report the GSE dynamical ranges spanned by definition (H, K, and M, respectively), while in the different coloured rectangles those of the other halo substructures are reported with the same legend as in the previous figures. The coloured points are the stars within each halo substructure sample that are chemically compatible with the pure GSE (H), again with the same colour legend. 
We can appreciate the very broad distribution of GSE in both angular momentum and energy, as expected from a massive merger (mass ratio $\ge$ 1:10) by simulations \citep[e.g., ][]{koppelman20, amarante22, khoperskov23b, khoperskov23d, mori24}.
We can then see that the compatible stars are actually sparsely distributed within each rectangle, not just concentrated in the closest regions to GSE. This fact means that the issue of associating the accreted stars to different merger debris is not simply due to an imprecise chosen value for the dynamical borders assumed to define the halo substructures. It rather hints that we have to rethink the necessity of defining new independent accretion events on the basis of kinematic cuts and chemical averages only. 

On the other hand, in the right panel of Figure \ref{fig:compatibility_elz}, we show the stars of each halo substructure that are chemically compatible with the metal-poor disc, again with the same colour legend for the data points and the defining rectangles. Unlike in the previous case, this time the distributions of chemically compatible stars are not spread out in a wide range of $E$ and $L_z$ within the dynamical definitions. 
We can see that the highest densities of disc-compatible stars are found in the most prograde and lowest energy regions of the rectangles, that is, the closest $E-L_z$ region to the disc's one, even though the compatibility was established on dynamically unbiased quantities (i.e., chemical abundances). However, this could be due to the intrinsic bias of the sample that features the highest density of data close to the dynamical distribution of the disc. 
Anyway, we can see that the substructures containing the highest percentages of stars chemically compatible with the metal-poor disc are either those on prograde orbits, such as the Helmi Stream (K) and (M), or those on low-energy retrograde or radial orbits, such as Thamnos, ad GSE (M), and Heracles. Substructures on retrograde and high-energy orbits, such as Sequoia, have a fraction of stars compatible with the disc which is null. This result is compatible with the prediction of numerical simulations, which show that the retrograde heated disc in the $E-L_z$ plane is usually confined to low-energy orbits, unless the early Milky Way had accreted a significant amount of mass through mergers \citep[i.e. about 50\% of its total mass at the time of accretion(s), see for example][]{jean-baptiste2017}.\\

In Table \ref{tab:sats}, we then summarise the fractions (in \%) of chemically compatible stars with respect to GSE (H) (first column) and the metal-poor disc (second column) for each of the halo substructures (different rows) when a threshold of 20 is assumed for the GMM analysis. In the third column, we also report the upper limits for their cleaned samples, computed taking into account that the GSE (H) itself features a 5\% chemical compatibility with the disc, that is subtracting this 5\% from the sum of GSE- and disc-compatible stars. 

\begin{figure}
   \centering
   \includegraphics[width=1\columnwidth]{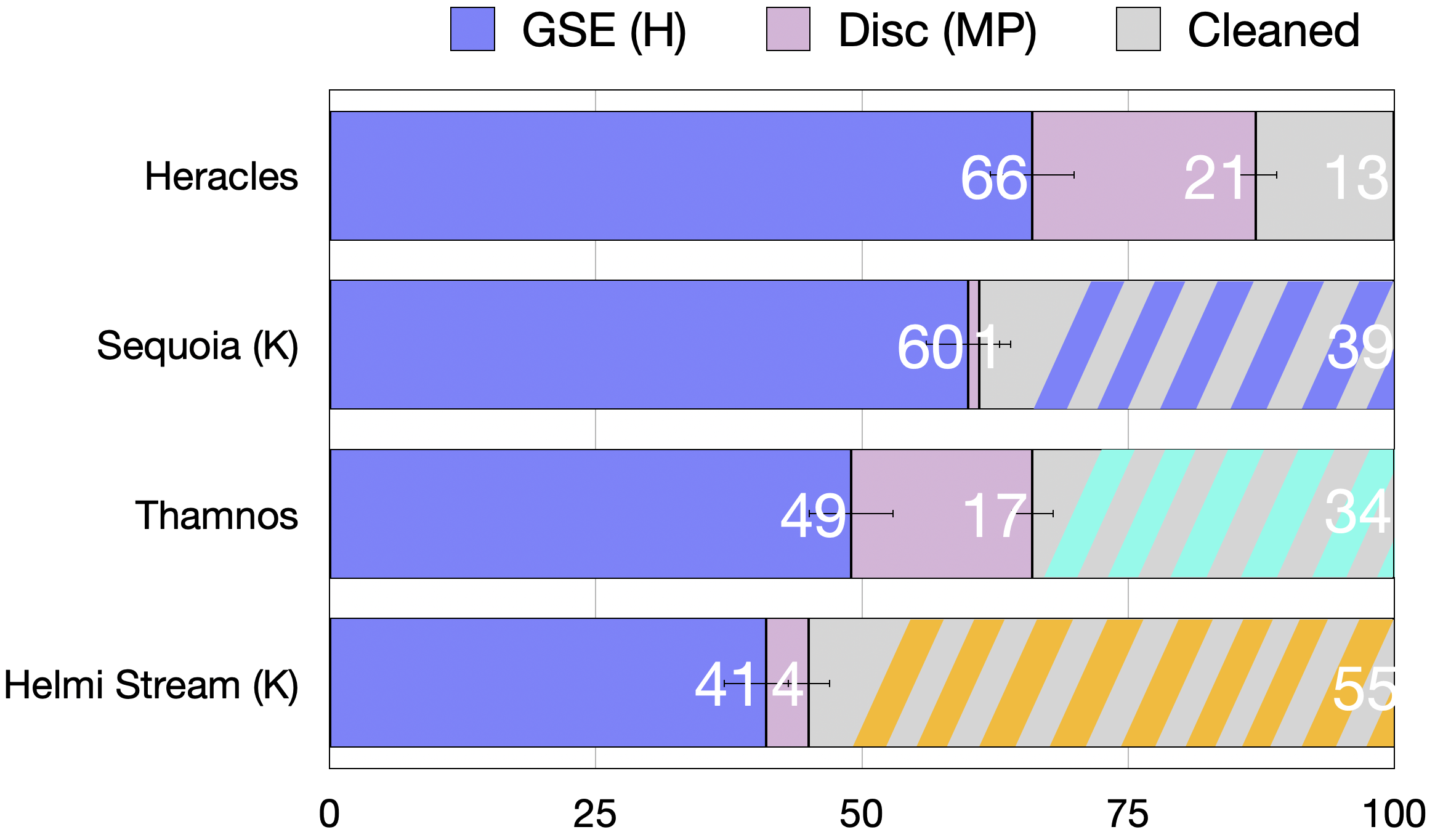}
   \caption{For each of the halo substructures listed to the left, we show the fractions (\%) of chemically compatible stars with respect to GSE (H) (in blue) and the metal-poor disc (in lilac), along with the lower limit for the cleaned samples (in grey), computed through the GMM analysis assuming a threshold of 20. The patterned grey highlights the similar chemical distribution of the cleaned Sequoia (K) to the GSE (H) outliers (blue), of the cleaned Thamnos to the distribution of $\omega$ Cen (turquoise), and of the cleaned Helmi Stream (K) to that of Sgr (orange).}
   \label{fig:comp_all}
\end{figure}

Moreover, in Figure \ref{fig:comp_all}, we summarise the fractions (in \%, for threshold=20) of chemically compatible stars with respect to GSE (H) (in blue) and the metal-poor disc (in lilac) for each of the halo substructures (listed to the left) for which we further analysed the chemical distributions of their cleaned samples. 
We recall the overlap in the chemical spaces of the cleaned Sequoia with the GSE (H) outliers (in blue), of the cleaned Thamnos with the distribution of $\omega$ Cen (in turquoise), and of the cleaned Helmi Stream (K) with that of Sgr (in orange). The similarity in their chemical distributions may hint at additional contamination from these systems.\\

\begin{table}[]
    \caption{Compatibility of halo substructures with GSE and the disc.}
    \centering
    \begin{tabular}{l | c c c }
        \hline
        \hline 
         & GSE (H) & Disc & Cleaned\\ 
        \hline
        Heracles          & 66 $\pm$ 5  & 21 $\pm$ 7 & $\le 18$ \\
        Sequoia (K)       & 60 $\pm$ 12 & 1  $\pm$ 3 & $\le 44$ \\
        Sequoia (My)      & 54 $\pm$ 6  & 1  $\pm$ 1 & $\le 50$ \\
        Sequoia (M)       & 48 $\pm$ 5  & 1  $\pm$ 1 & $\le 56$ \\
        Thamnos           & 49 $\pm$ 6  & 17 $\pm$ 6 & $\le 39$ \\
        Helmi Stream (K)  & 41 $\pm$ 7  & 4  $\pm$ 3 & $\le 60$ \\
        Helmi Stream (M)  & 52 $\pm$ 4  & 12 $\pm$ 2 & $\le 41$ \\
        Sagittarius       & 31 $\pm$ 7  & 0  $\pm$ 0 & $\le 74$ \\
        $\omega$ Centauri & 2  $\pm$ 1  & 0  $\pm$ 0 & $\le 100$ \\
        \hline 
    \end{tabular}
    \tablefoot{The fractions (expressed in \%) of GSE/disc chemically compatible stars are computed assuming for the GMM a threshold of 20. The GSE definition is the H23 one, while for the disc it is the metal-poor sample. We remind that the fractions are relative to the threshold assumed (20) and must be renormalised with the maximum possible (80) for absolute values. The cleaned samples' fractions are upper limits, taking into account that the GSE (H) itself features a 5\% chemical compatibility with the disc.}
    \label{tab:sats}
\end{table}

We emphasise that the chemical compatibility measured in the previous sections for the different halo substructures cannot be directly interpreted as a common origin between different substructures. If we consider Heracles and Sequoia, for example, they both have a large fraction of stars compatible with GSE, however the bulk of Heracles stars compatible with GSE has [Fe/H] values below $-1$, while GSE extends well above this threshold. If this chemical compatibility points to a common origin, this behaviour may be explained by a repartition of accreted GSE stars in the $E-Lz$ plane which depends on metallicity, which would be a natural outcome if the GSE progenitor had an initial metallicity gradient. However, this shift in the mean metallicity of GSE and GSE-compatible stars in Heracles may also point out an intrinsically different origin of these two samples of stars. \\
As for Sequoia, the high level of compatibility that this substructure shows with GSE is also accompanied by a more homogeneous distribution of its stars over the whole range of chemical abundances spanned by GSE -- making the genetic link between these two samples more sound than for the Heracles-GSE case. Also for Sequoia, however, we have found that there may be some contamination -- at a few percent level -- of relatively metal-rich stars (i.e. with [Fe/H]$\gtrsim -1$) which have lower $\alpha$-abundances than GSE stars at similar metallicities. \\
While Heracles possibly shares a common origin both with the GSE and with the disc populations, and Sequoia is plausibly associated to GSE, the Helmi Stream is even a more complex mixture of stars. Once cleaned of the possible GSE and heated disc contaminants, the remaining stars show a striking overlap in chemical abundances with the Sagittarius dwarf. This is interesting, given also its proximity in energy and angular momentum. However, this result opens the question about the real nature of this system, whether there is an intrinsic and specific signature of an accretion event (visible in chemical abundances) or whether this structure is the overlap of stars having different origins. It also invites to reconsider the extension in the $E-L_z$ space of stars lost by the Sagittarius dwarf, part of which seems to fall in the Helmi stream selection, part in the GSE selection itself. \\

Finally, this work used the same dataset and the same selections (for most substructures) as those used by \citet{horta23}.
Also in H23, the objective was to compare substructures defined on the basis of mainly kinematic criteria in order to quantify their similarity in chemical spaces. There are some methodological differences between the two works, however. 
The way of quantifying the chemical similarities or differences between substructures is not the same. We take into account all the stars that make up a substructure and quantify the fraction of stars compatible with another substructure, through a GMM approach. H23, on the other hand, use a restricted subset of stars from a given substructure, all with the same specific metallicity -- which may vary depending on the substructure considered. For these stars, they estimate the average abundances (relative to iron), which they then compare with those of a comparison substructure. H23's approach, by its very nature, does not allow to determine whether a structure could be a mixture of several others. \\ 
These different approaches, and the impossibility to capture contamination by using mean values,  lead H23 to conclude, for example,  that GSE and the Helmi stream have a null probability to be chemically compatible, while -- as discussed in the previous section -- we can assess that in fact  40\% of stars in the Helmi stream are chemically compatible with GSE, concluding that this stream of stars contains also about 5\% of stars chemically compatible with the disc and the remaining compatible with the chemical abundances of Sagittarius dwarf stars. Another example is given by Heracles, for which a chemical compatibility with both GSE and the disc is excluded by H23. As we have seen in the previous sections, our analysis leads to a different result, suggesting that about 65\% of Heracles stars are chemically compatible with GSE and 20\% with the metal-poor disc. 
These few examples demonstrate how subtle the analysis of these structures can be, and also the complexity that can be hidden in these "systems", which — it is worth remembering — are often defined on the basis of kinematic criteria alone, the limitations of which have been demonstrated in recent years (see introduction).

Once it has been clarified that regions in the $E-L_z$ space cannot be studied as single populations, making use of their chemical abundances to disentangle their origin becomes much more complex.

\section{Conclusions}
\label{sec:conclusions}

In this work, we took advantage of the \textit{Gaia} EDR3 and APOGEE DR17 catalogues to compare from a chemical point of view the halo substructures of the Milky Way that are (quasi-only) defined with dynamical cuts. The objective of the study was to compare the minor accretions of the MW to the last massive merger event of the MW (i.e., GSE) and to the in situ component (that is, the heated disc), in order to clean their samples from these two sources of contamination and study their intrinsic chemical patterns. 
To this aim, we firstly built a Gaussian Mixture Model assuming a threshold of the 20th percentile of the GSE distribution in a 7-dimensional chemical space given by the following abundances: [Fe/H], [Mg/Fe], [Si/Fe], [Ca/Fe, [C/Fe], [Mn/Fe], and [Al/Fe]. These elements allow a thorough analysis of different nucleosynthetic channels ($\alpha$, iron peak, odd-Z light elements) to probe different chemical enrichment histories on a star-by-star basis, while preserving a very high-precision sample (median uncertainties $\le 0.02$). 
Given the assumed threshold, we found that:
\begin{itemize}
    \item A number of halo substructures (e.g., Heracles, Sequoia, and Thamnos) contain a significant ($\gtrsim 50\%$) percentage of stars chemically compatible with GSE. Among them, Sequoia is the one for which the chemical compatibility with GSE seems to robustly indicate a common origin, that is both a high compatibility fraction (66\%) and overall overlap of chemical distributions. 
    \item Other systems, such as the Helmi stream and Sagittarius, show an intermediate level (between 30\% and 40\%) of compatibility with GSE. 
    \item $\omega$ Cen features a negligible (2\%) chemical compatibility with GSE, although commonly claimed to be its nuclear star cluster in the literature. 
    \item Around half of the samples analysed also feature non-negligible contamination from the metal-poor disc ($\sim$20\%). In particular, substructures on prograde orbits, as the Helmi Stream, but also on retrograde/radial orbits with low energy, as Thamnos and Heracles, are found to be more contaminated by the metal-poor disc. The latter are also $\alpha$-enhanced, highlighting a correlation between the "high-$\alpha$" substructures and the disc contamination.
    \item The GSE compatible stars are spread throughout the $E-L_z$ space, but also within the entire kinematic regions of the substructures.
    This calls into question the true nature and physical significance of these substructures.
\end{itemize}

Once the contamination by GSE and the heated disc have been removed, the cleaned substructures still reveal a number of interesting features, which lead us to suggest the following:
\begin{itemize}
\item Sequoia, while heavily dominated by GSE stars, contains a small fraction of stars (between 5 and 10\%) which seem to deviate from the GSE chemical abundance trends, these stars being less $\alpha$-enhanced (in Mg and Si) than GSE stars at the same metallicities ([Fe/H]$\gtrsim -1$). They could represent the less efficiently chemically enriched outskirts of the GSE original galaxy.
\item Thamnos cleaned sample ($\sim$35\%) is mostly made of stars whose chemical patterns overlap with those of $\omega$~Cen. This similarity is interesting, given also the proximity of Thamnos and $\omega$ Cen in the $E-L_z$ space, possibly suggesting an interconnection between the two systems. 
\item While half of the Helmi stream appears to be chemically compatible either with GSE ($\sim$40\% of its stars) or with the metal-poor disc ($\sim$5\% of its stars), the remaining half shows, for the large majority, an impressive similarity to the chemical patterns of the Sagittarius dSph and its stream. Once more, this fact together both with the proximity in the $E-L_z$ and $L_\perp-L_z$ spaces and in their spatial distribution, hints a non-negligible contamination from Sgr within the Helmi Stream. 
\item The significant compatibility of the chemical patterns of Sagittarius with those of GSE ($\sim$30\%) seems to suggest that stars from the Sagittarius stream do contaminate the GSE sample as well, a fortiori given the overlap between the two substructures in the $E-L_z$ space at high energies. 
\end{itemize}

In summary, none of these substructures appears as a unique population of stars with its own origin, all show some chemical overlap with GSE, the metal-poor disc, Sagittarius and even $\omega$~Cen. The GSE region itself possibly contains part of the Sagittarius stream.

\begin{acknowledgements}

AM and SS acknowledge support from the ERC Starting Grant NEFERTITI H2020/808240. 
MH and PDM acknowledge  the support of the French Agence Nationale de la Recherche (ANR), under grant ANR-13-BS01-0005 (project ANR-20-CE31-0004-01 MWDisc). 
This work has made use of data from the European Space Agency (ESA) mission Gaia (https://www.cosmos.esa.int/gaia), processed by the Gaia Data Processing and Analysis Consortium (DPAC, https://www.cosmos.esa.int/web/gaia/dpac/consortium). Funding for the DPAC has been provided by national institutions, in particular the institutions participating in the Gaia Multilateral Agreement. 
This study makes use of the astroNN catalog (Leung \& Bovy 2019, https://github.com/henrysky/astroNN). 
Funding for the Sloan Digital Sky Survey IV has been provided by the Alfred P. Sloan Foundation, the U.S. Department of Energy Office of Science, and the Participating Institutions. SDSS-IV acknowledges support and resources from the Center for High Performance Computing at the University of Utah. The SDSS website is www.sdss.org. SDSS-IV is managed by the Astrophysical Research Consortium for the Participating Institutions of the SDSS Collaboration including the Brazilian Participation Group, the Carnegie Institution for Science, Carnegie Mellon University, Center for
Astrophysics | Harvard \& Smithsonian, the Chilean Participation Group, the French Participation Group, Instituto de Astrofísica de Canarias, The Johns
Hopkins University, Kavli Institute for the Physics and Mathematics of the Universe (IPMU)/University of Tokyo, the Korean Participation Group, Lawrence Berkeley National Laboratory, Leibniz Institut für Astrophysik Potsdam (AIP), Max-Planck-Institut für Astronomie (MPIA Heidelberg), Max-Planck-Institut für Astrophysik (MPA Garching), Max-Planck-Institut für Extraterrestrische Physik (MPE), National Astronomical Observatories of China, New Mexico State University, New York University, University of Notre Dame, Observatário Nacional/MCTI, The Ohio State University, Pennsylvania State University, Shanghai Astronomical Observatory, United Kingdom Participation Group, Universidad Nacional Autónoma de México, University of Arizona, University of Colorado Boulder, University of Oxford, University of Portsmouth, University of Utah, University of Virginia, University of Washington, University of Wisconsin,
Vanderbilt University, and Yale University.

\end{acknowledgements}

\bibliographystyle{aa}
\bibliography{biblio} 

\clearpage
\begin{appendix}
\clearpage

\section{Chemical compatibility percentages of the halo substructures with GSE (H) and the metal-poor disc}
\label{appendix:tables}

\begin{table*}[]
    \caption{Fraction of stars chemically compatible with GSE.}
    \label{tab:comp}
    \centering
    \resizebox{\linewidth}{!}{
    \begin{tabular}{c c c c c c c c c c}
        \hline\hline 
            & Heracles & Sequoia (K) & Sequoia (My) & Sequoia (M) & Thamnos & Helmi (K) & Helmi (M) & Sgr & $\omega$ Cen\\ 
        $N_{*}$ & 222 & 18 & 107 & 131 & 118 & 54 & 427 & 263 & 400\\
        \hline\hline 
        \multicolumn{9}{c}{GSE (H)}\\
        \hline         
            $f_{comp}(th=5)$  (\%) & 90 $\pm$ 3 & 84 $\pm$ 10 & 78 $\pm$ 5 & 73 $\pm$ 5 & 81 $\pm$ 5 & 75 $\pm$ 8 & 80 $\pm$ 4 & 68 $\pm$ 9  & 21 $\pm$ 4 \\
            $f_{comp}(th=10)$ (\%) & 82 $\pm$ 4 & 73 $\pm$ 11 & 68 $\pm$ 5 & 62 $\pm$ 5 & 69 $\pm$ 5 & 59 $\pm$ 9 & 68 $\pm$ 4 & 52 $\pm$ 10 & 10 $\pm$ 3 \\ 
            $f_{comp}(th=15)$ (\%) & 74 $\pm$ 4 & 67 $\pm$ 12 & 60 $\pm$ 5 & 54 $\pm$ 5 & 58 $\pm$ 6 & 48 $\pm$ 8 & 60 $\pm$ 4 & 40 $\pm$ 8  & 5  $\pm$ 2 \\
            $f_{comp}(th=20)$ (\%) & 66 $\pm$ 5 & 60 $\pm$ 12 & 54 $\pm$ 6 & 48 $\pm$ 5 & 49 $\pm$ 6 & 41 $\pm$ 7 & 52 $\pm$ 4 & 31 $\pm$ 7  & 2  $\pm$ 1 \\ 
        \hline\hline
        \multicolumn{9}{c}{GSE (K)}\\ 
        \hline         
            $f_{comp}(th=5)$  (\%) & 85 $\pm$ 4 & 79 $\pm$ 11 & 75 $\pm$ 6 & 71 $\pm$ 6 & 70 $\pm$ 7 & 69 $\pm$ 9 & 68 $\pm$ 8 & 69 $\pm$ 18 & 16 $\pm$ 5 \\ 
            $f_{comp}(th=10)$ (\%) & 76 $\pm$ 5 & 70 $\pm$ 12 & 65 $\pm$ 6 & 61 $\pm$ 6 & 57 $\pm$ 7 & 56 $\pm$ 9 & 55 $\pm$ 8 & 53 $\pm$ 18 & 8  $\pm$ 3 \\ 
            $f_{comp}(th=15)$ (\%) & 68 $\pm$ 6 & 64 $\pm$ 12 & 58 $\pm$ 6 & 55 $\pm$ 6 & 48 $\pm$ 7 & 47 $\pm$ 9 & 46 $\pm$ 8 & 43 $\pm$ 17 & 4  $\pm$ 2 \\
            $f_{comp}(th=20)$ (\%) & 61 $\pm$ 6 & 58 $\pm$ 13 & 52 $\pm$ 6 & 49 $\pm$ 6 & 40 $\pm$ 7 & 39 $\pm$ 8 & 37 $\pm$ 7 & 35 $\pm$ 14 & 2  $\pm$ 1 \\ 
        \hline\hline
        \multicolumn{9}{c}{GSE (M)}\\
        \hline         
            $f_{comp}(th=5)$  (\%) & 91 $\pm$ 2 & 82 $\pm$ 10 & 77 $\pm$ 4 & 70 $\pm$ 5 & 89 $\pm$ 3 & 75 $\pm$ 7 & 84 $\pm$ 2 & 68 $\pm$ 6 & 23 $\pm$ 3 \\
            $f_{comp}(th=10)$ (\%) & 83 $\pm$ 3 & 72 $\pm$ 11 & 66 $\pm$ 5 & 59 $\pm$ 5 & 80 $\pm$ 4 & 59 $\pm$ 8 & 74 $\pm$ 3 & 52 $\pm$ 7 & 11 $\pm$ 2 \\
            $f_{comp}(th=15)$ (\%) & 75 $\pm$ 3 & 64 $\pm$ 12 & 57 $\pm$ 5 & 50 $\pm$ 5 & 70 $\pm$ 5 & 48 $\pm$ 8 & 67 $\pm$ 3 & 39 $\pm$ 7 & 5  $\pm$ 2 \\ 
            $f_{comp}(th=20)$ (\%) & 67 $\pm$ 4 & 56 $\pm$ 13 & 49 $\pm$ 5 & 43 $\pm$ 5 & 60 $\pm$ 5 & 39 $\pm$ 7 & 60 $\pm$ 3 & 30 $\pm$ 6 & 2  $\pm$ 1 \\
        \hline
    \end{tabular}
    }
    \tablefoot{Total number of stars and fraction of chemically compatible stars (with relative uncertainties) for MW halo substructure (different columns) with respect to the different GSE definitions (from top to bottom, GSE (H), GSE (K), and GSE (M)), assuming different thresholds (different rows), i.e. increasing purity of the sample from threshold 5\% to 20\%. We remind that the fractions are relative to the assumed threshold and must be renormalised with the maximum possible (95, 90, 85, 80 when assuming thresholds=5, 10, 15, 20) for absolute values.}
\end{table*}

\begin{table*}[]
    \caption{Fraction of stars chemically compatible with the disc. }
    \label{tab:comp_disc}
    \centering
    \resizebox{\linewidth}{!}{
    \begin{tabular}{c c c c c c c c c c c c c c}
        \hline\hline 
            & GSE (H) & GSE (K) & GSE (M) & Heracles & Seq (K) & Seq (My) & Seq (M) & Helmi (K) & Helmi (M) & Thamnos & Sgr & $\omega$ Cen\\
        $N_{*}$ & 941 & 442 & 3332 & 222 & 18 & 107 & 131 & 54 & 427 & 118 & 263 & 400\\
        \hline\hline 
        \multicolumn{11}{c}{Metal poor disc} \\
        \hline
            $f_{comp}(th=5)$ (\%)  & 39 $\pm$ 9& 37 $\pm$ 10& 43 $\pm$ 7& 72 $\pm$ 7& 21 $\pm$ 13& 24 $\pm$ 9& 19 $\pm$ 7& 16 $\pm$ 6& 32 $\pm$ 6& 62 $\pm$ 8& 1 $\pm$ 1& 5  $\pm$ 2\\
            $f_{comp}(th=10)$ (\%) & 19 $\pm$ 6& 18 $\pm$ 6& 28 $\pm$ 4& 52 $\pm$ 7& 8 $\pm$ 8& 8 $\pm$ 4& 9 $\pm$ 5& 9 $\pm$ 4& 21 $\pm$ 3& 42 $\pm$ 8& 0 $\pm$ 00& 1 $\pm$ 1\\
            $f_{comp}(th=15)$ (\%) & 10 $\pm$ 3& 8 $\pm$ 3& 20 $\pm$ 3& 35 $\pm$ 7& 3  $\pm$ 5& 3  $\pm$ 2& 4  $\pm$ 2& 6 $\pm$ 3& 16 $\pm$ 3& 28 $\pm$ 8& 0  $\pm$ 0& 0  $\pm$ 0\\
            $f_{comp}(th=20)$ (\%) & 5  $\pm$ 2& 4  $\pm$ 2& 15 $\pm$ 2& 21 $\pm$ 7& 1  $\pm$ 3& 1  $\pm$ 1& 1  $\pm$ 1& 4  $\pm$ 3& 12 $\pm$ 2& 17 $\pm$ 6& 0  $\pm$ 0& 0  $\pm$ 0\\
        \hline
    \end{tabular}
    }
    \tablefoot{Total number of stars and fraction of chemically compatible stars (with relative uncertainties) for every MW halo substructure (different columns) with respect to the metal-poor disc ($\rm{[Fe/H]} \le -0.8$), assuming different thresholds (different rows), i.e. increasing purity of the sample from threshold 5\% to 20\%. We remind that the fractions are relative to the threshold assumed and must be renormalised with the maximum possible (95, 90, 85, 80 for thresholds=5, 10, 15, 20) for absolute values.}
\end{table*}

The fractions of stars within each halo substructure chemically compatible with respect to GSE are reported in the different columns of Table \ref{tab:comp}, along with the total number of stars and the associated uncertainty, computed by bootstrap resampling the data ($n_{boot}= 1000$) as described in Section \ref{sec:method}. 
Furthermore, we list the results obtained considering different threshold values for the accepted probability density, which are 5, 10, 15 and 20 percentiles, i.e. increasing purity of the sample from threshold 5\% to 20\%. The fractions vary accordingly, but the ranking of the most chemically-compatible substructures is not affected. 
Finally, we show the chemical comparison with respect to the three different definitions of GSE considered in this work (H, K, and M), the results of which are generally consistent with each other within the uncertainties.

For each halo substructure, in the different columns of Table \ref{tab:comp_disc} we report the fractions of stars chemically compatible with the metal-poor disc along with the total number of stars and their associated uncertainty ($n_{boot}= 1000$). 
As for the case of the comparison with GSE (Table~\ref{tab:comp}), we list the results obtained considering different threshold values for the accepted probability density, that are 5, 10, 15 and 20 percentiles, i.e. increasing purity of the sample from threshold 5\% to 20\%. The fractions vary accordingly, but the following results hold true.


\section{Chemical spaces comparison of the halo substructures to GSE (H) and the metal-poor disc}
\label{appendix:comp}

\begin{figure*}
    \centering
    \includegraphics[width=1.75\columnwidth]{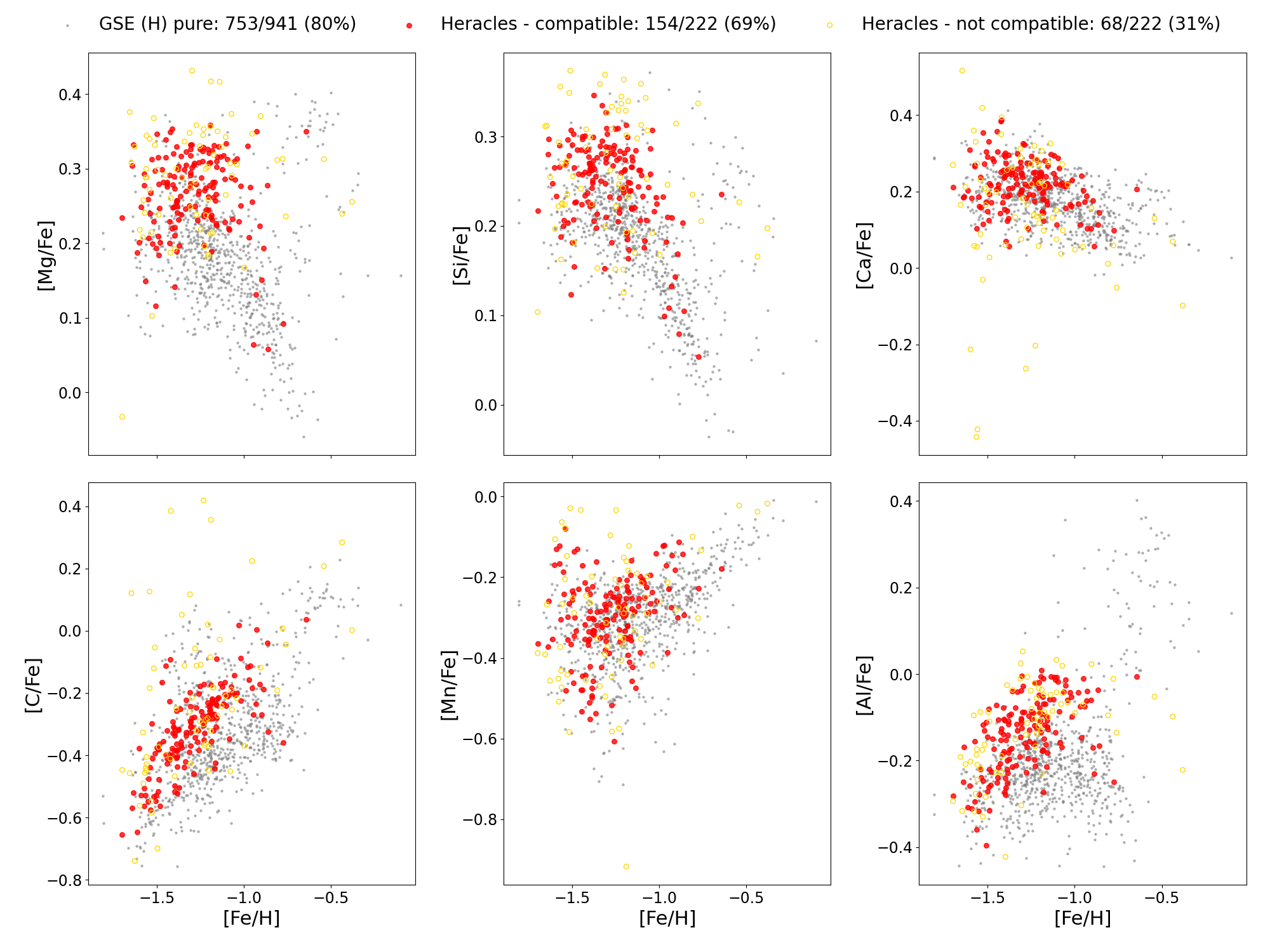}
    \includegraphics[width=1.75\columnwidth]{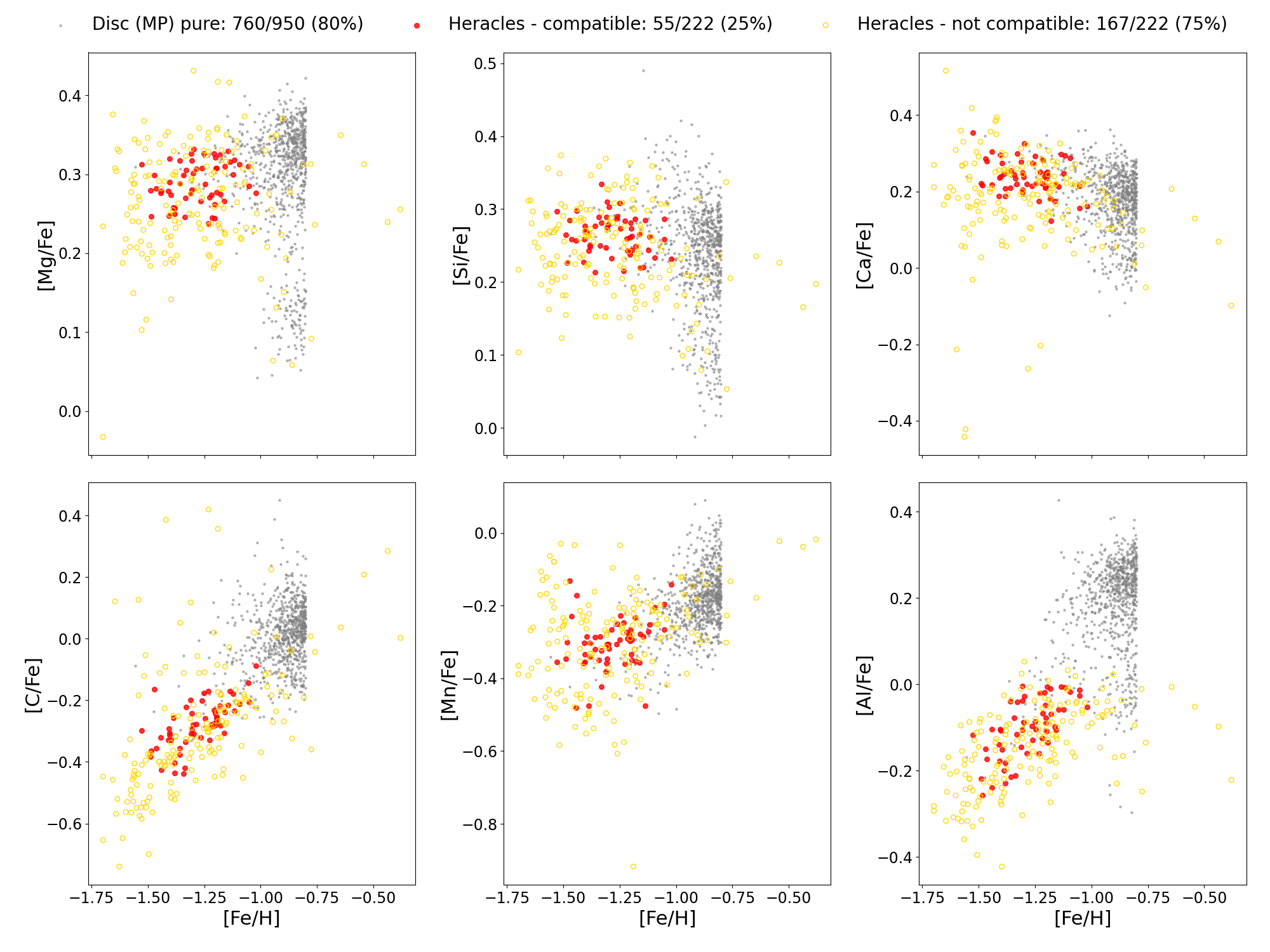}
    \caption{Heracles' stars that are chemically compatible (threshold=20) with GSE (H) (upper panel) and with the metal-poor disc (lower panel) are shown in red, while the ones that are not compatible in yellow. The relative percentages are reported in the legend (and have to be normalised to 80\% for absolute values). In grey we show the bulk (80\%) of GSE and of the metal-poor disc, in the upper and lower panels, respectively. }
    \label{fig:heracles_cut}
\end{figure*}

\begin{figure*}
    \centering
    \includegraphics[width=1.75\columnwidth]{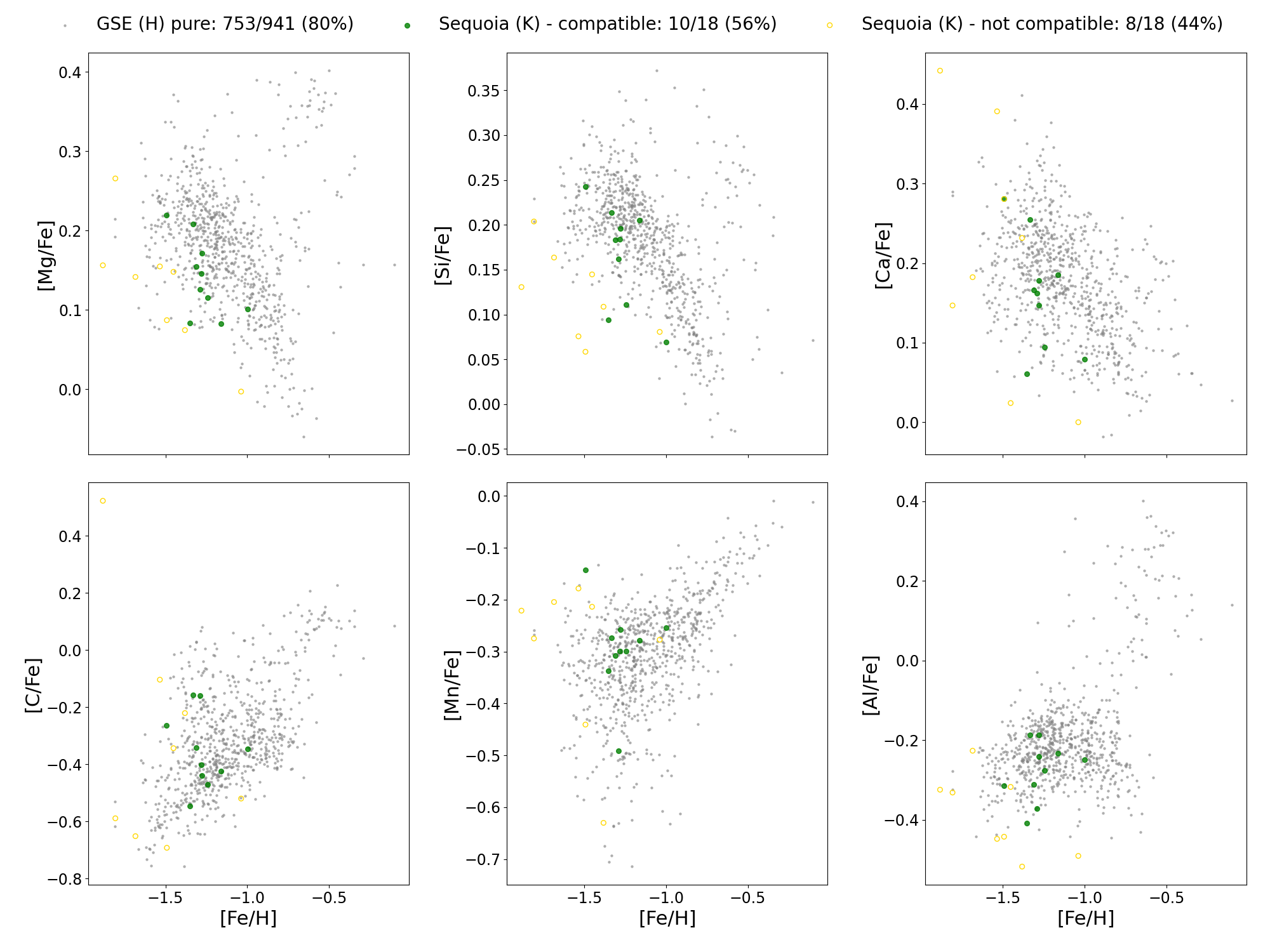}
    \includegraphics[width=1.75\columnwidth]{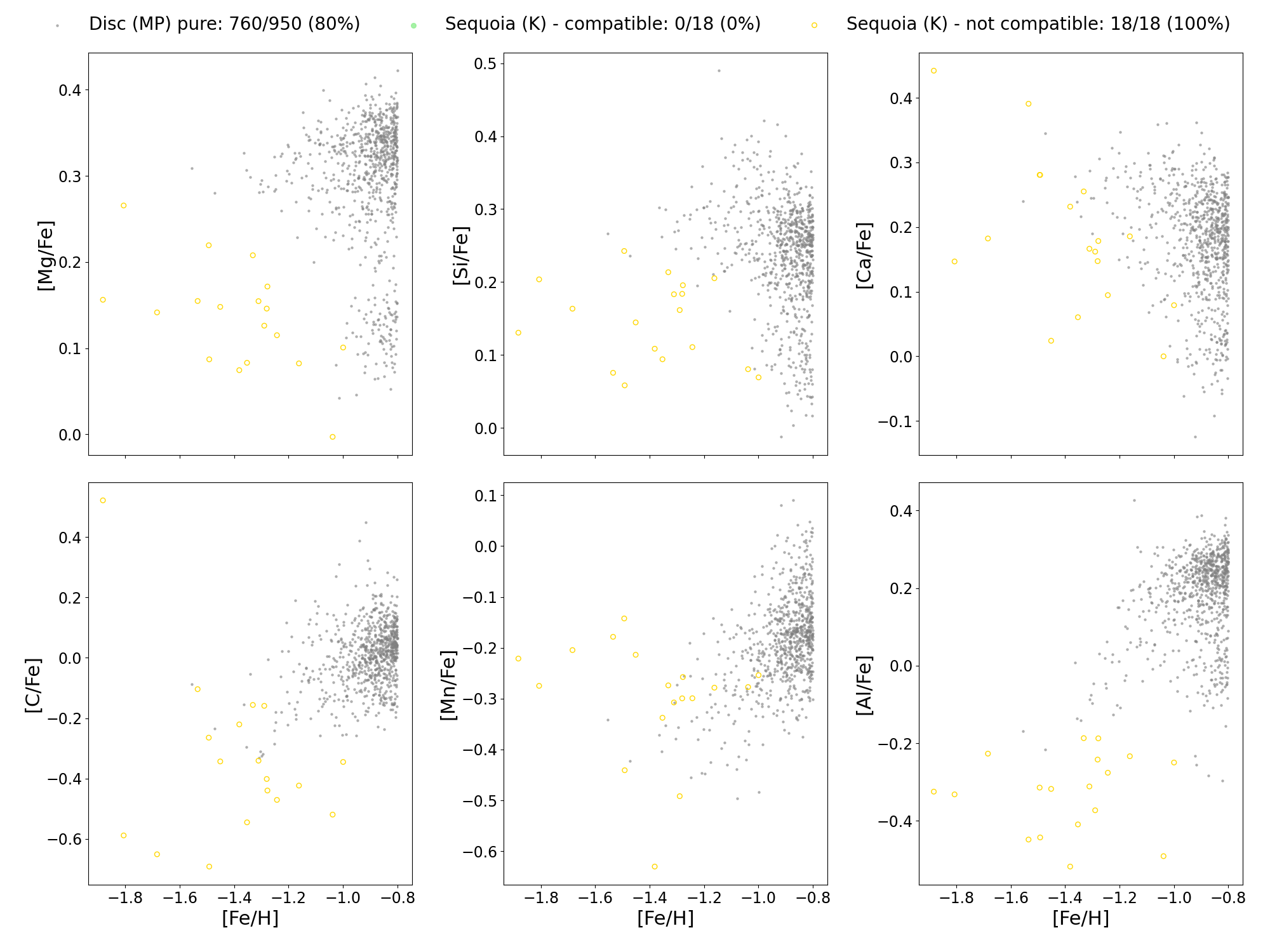}
    \caption{Sequoia (K)'s stars that are chemically compatible (threshold=20) with GSE (H) (upper panel) and with the metal-poor disc (lower panel) are shown in green, while the ones that are not compatible in yellow. The relative percentages are reported in the legend (and have to be normalised to 80\% for absolute values). In grey we show the bulk (80\%) of GSE and of the metal-poor disc, in the upper and lower panels, respectively. }
    \label{fig:sequoia_cut}
\end{figure*}

\begin{figure*}
    \centering
    \includegraphics[width=1.75\columnwidth]{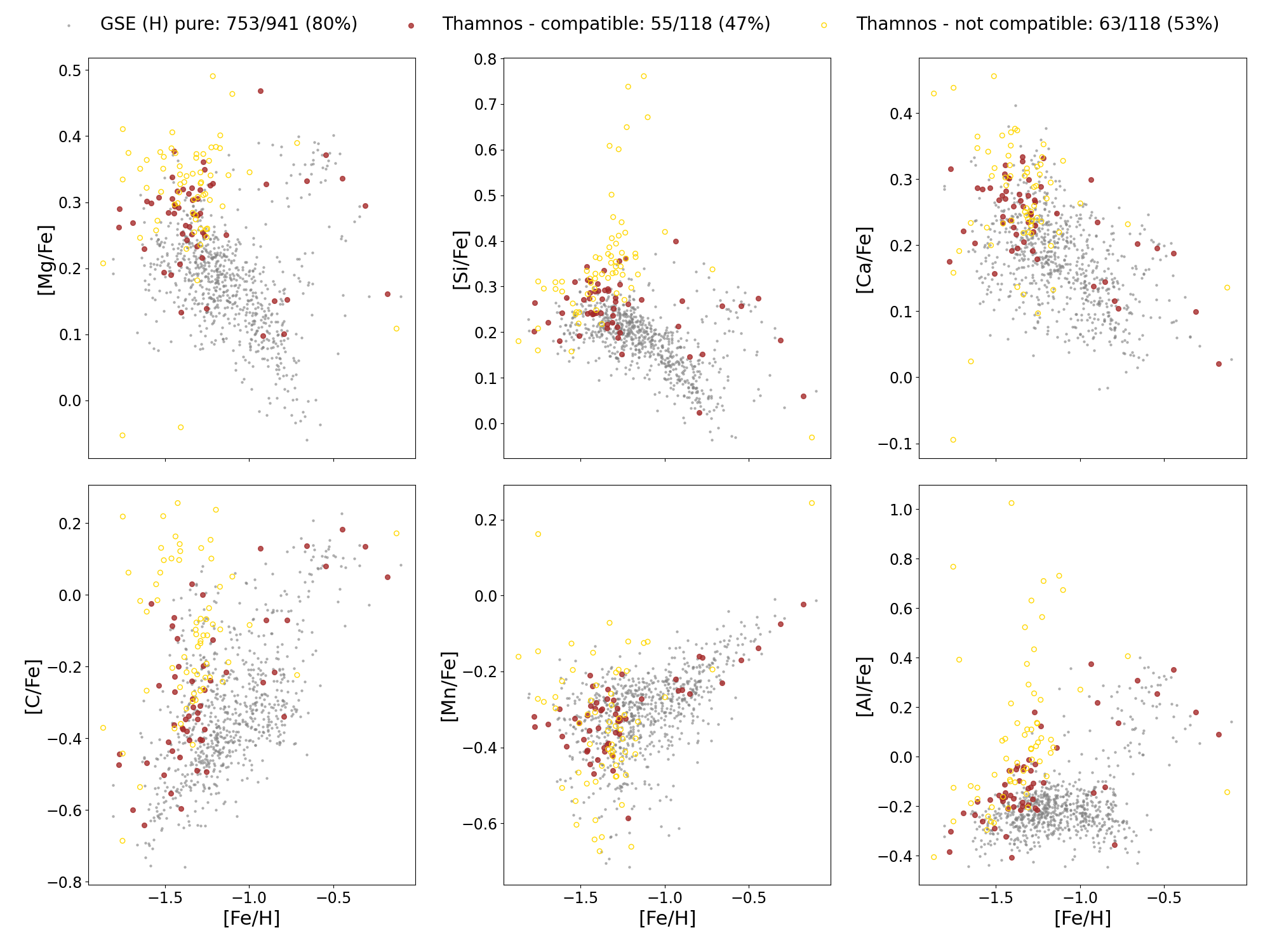}
    \includegraphics[width=1.75\columnwidth]{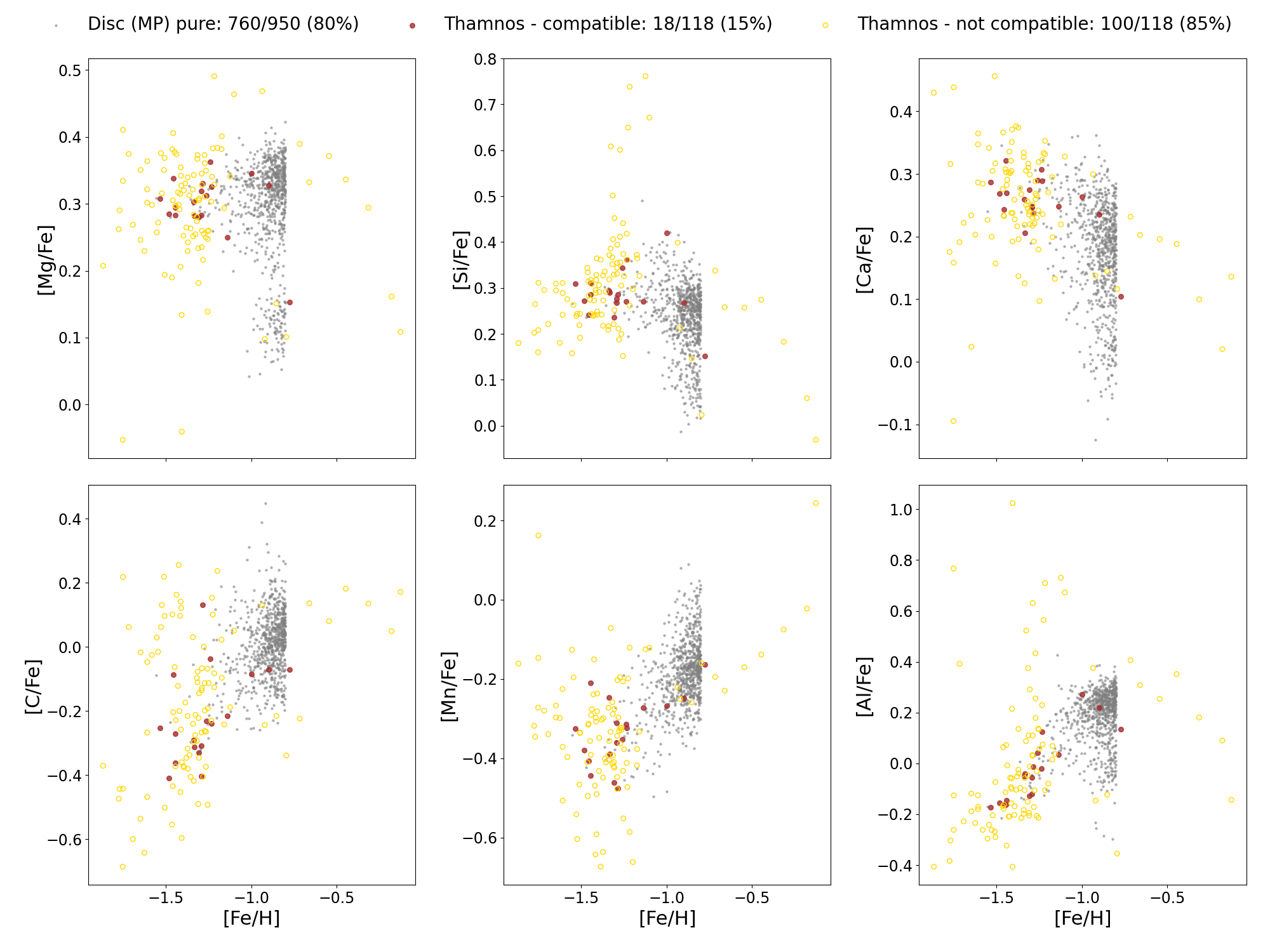}
    \caption{Thamnos' stars that are chemically compatible (threshold=20) with GSE (H) (upper panel) and with the metal-poor disc (lower panel) are shown in brown, while the ones that are not compatible in yellow. The relative percentages are reported in the legend (and have to be normalised to 80\% for absolute values). In grey we show the bulk (80\%) of GSE and of the metal-poor disc, in the upper and lower panels, respectively.
    } 
    \label{fig:thamnos_cut}
\end{figure*}

\begin{figure*}
    \centering
    \includegraphics[width=1.75\columnwidth]{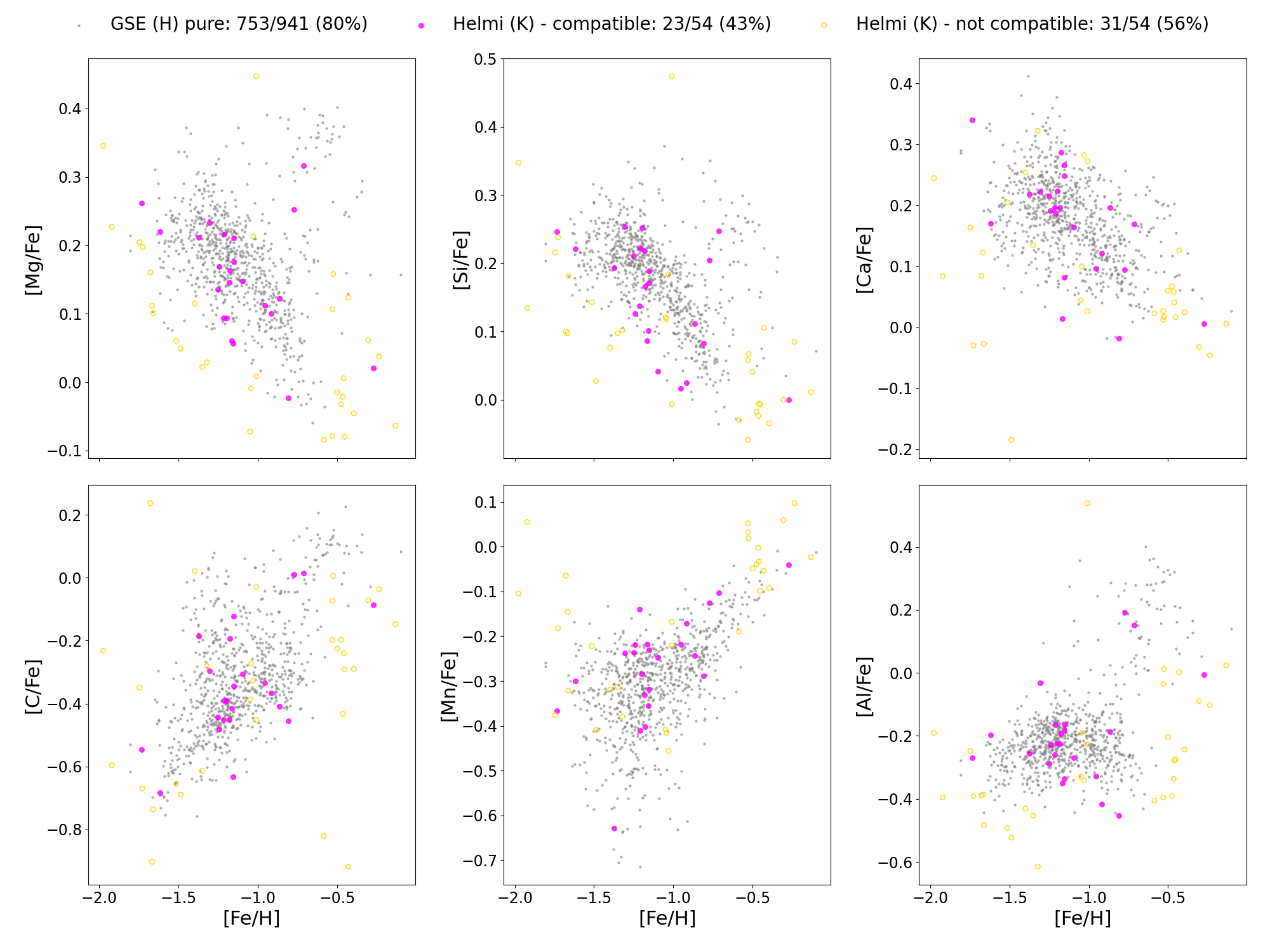}
    \includegraphics[width=1.75\columnwidth]{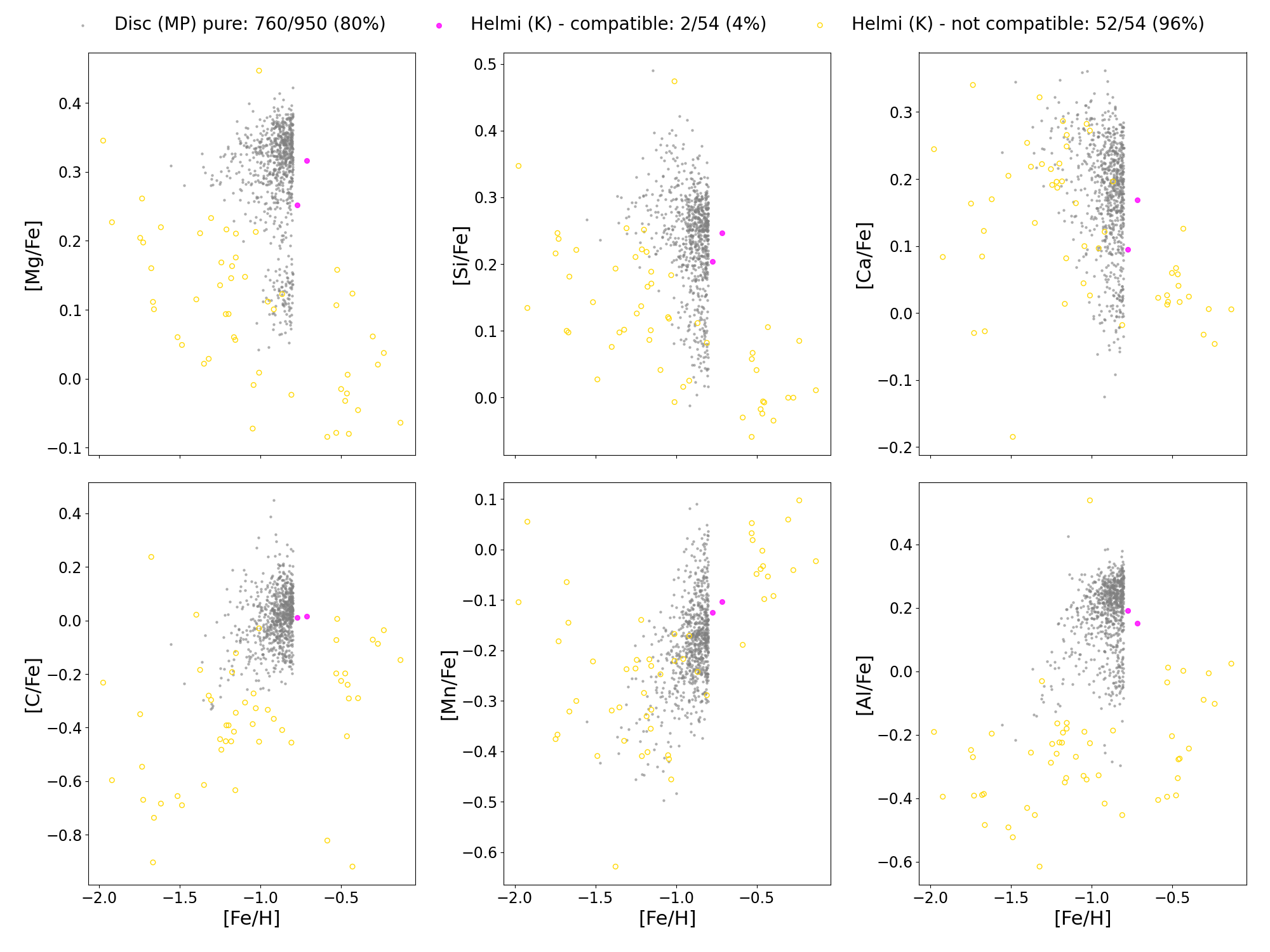}
    \caption{Helmi Stream (K)'s stars that are chemically compatible (threshold=20) with GSE (H) (upper panel) and with the metal-poor disc (lower panel) are shown in magenta, while the ones that are not in yellow. The relative percentages are reported in the legend (and have to be normalised to 80\% for absolute values). In grey we show the bulk (80\%) of GSE and of the metal-poor disc, in the upper and lower panels, respectively.}
    \label{fig:helmi_cut}
\end{figure*}

\begin{figure*}
    \centering
    \includegraphics[width=1.75\columnwidth]{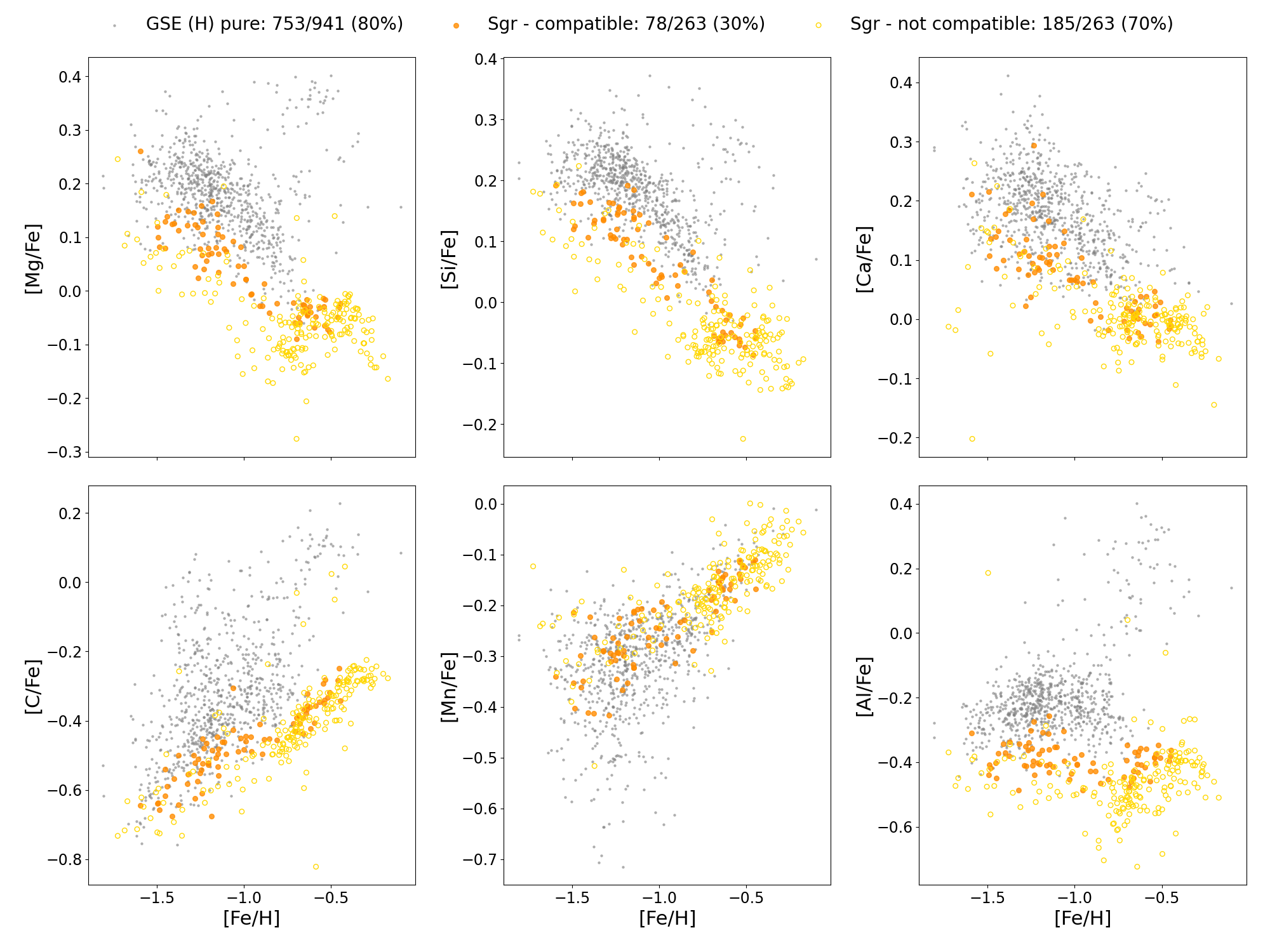}
    \includegraphics[width=1.75\columnwidth]{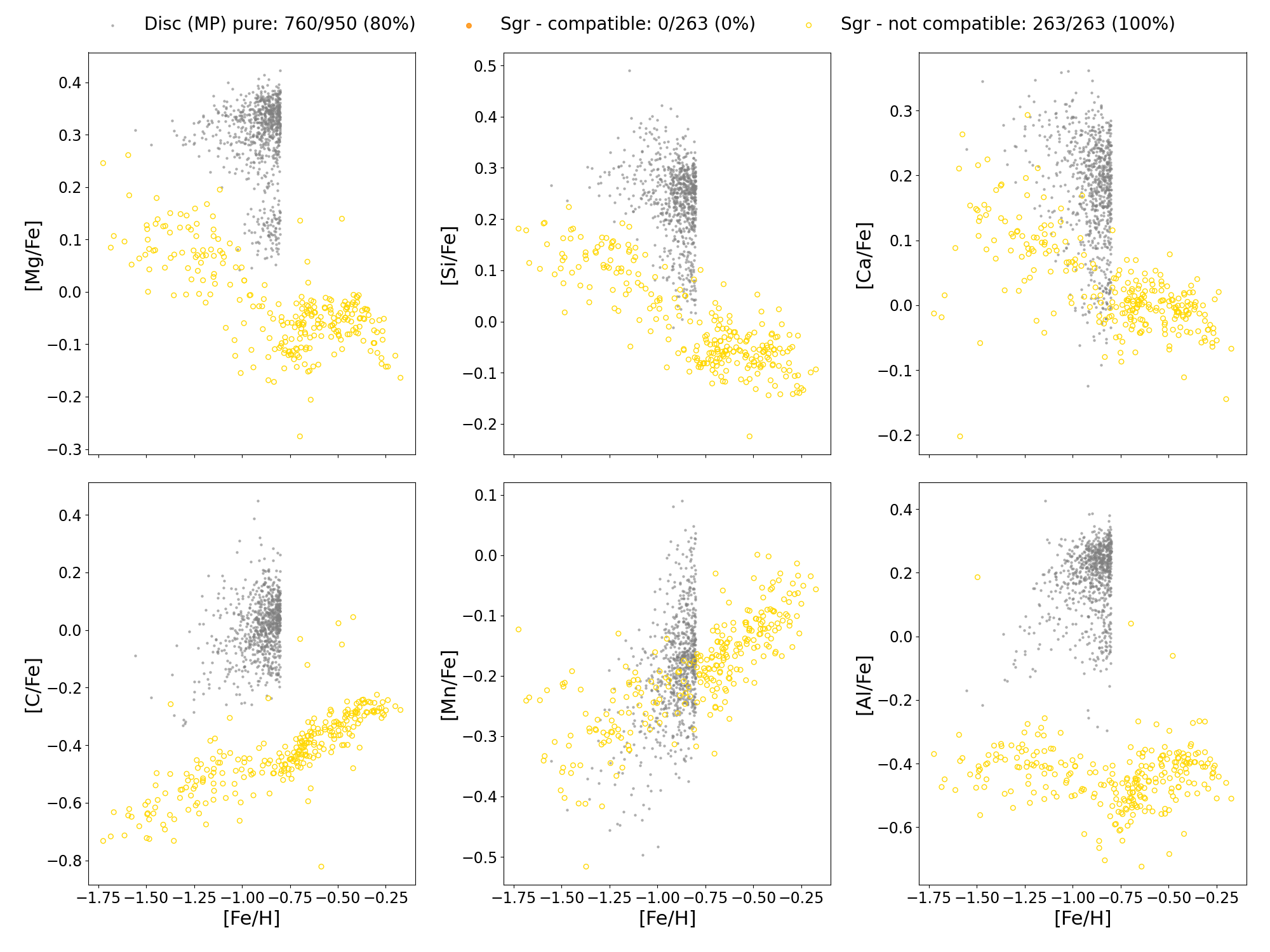}
    \caption{Sagittarius' stars that are chemically compatible (threshold=20) with GSE (H) (upper panel) and with the metal-poor disc (lower panel) are shown in orange, while the ones that are not compatible in yellow. The relative percentages are reported in the legend (and have to be normalised to 80\% for absolute values). In grey we show the bulk (80\%) of GSE and of the metal-poor disc, in the upper and lower panels, respectively.}
    \label{fig:sgr_cut}
\end{figure*}

\begin{figure*}
    \centering
    \includegraphics[width=1.75\columnwidth]{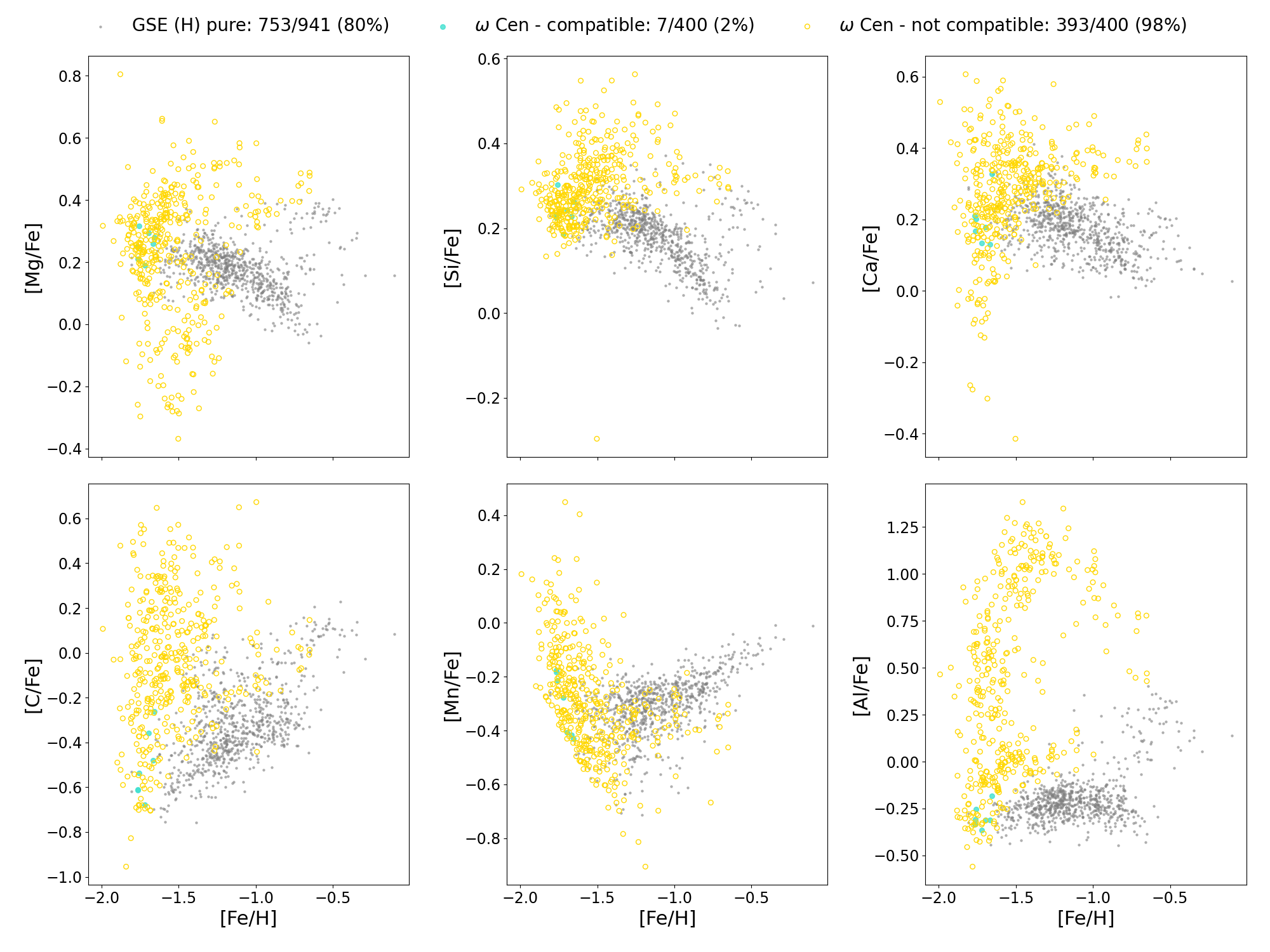}
    \includegraphics[width=1.75\columnwidth]{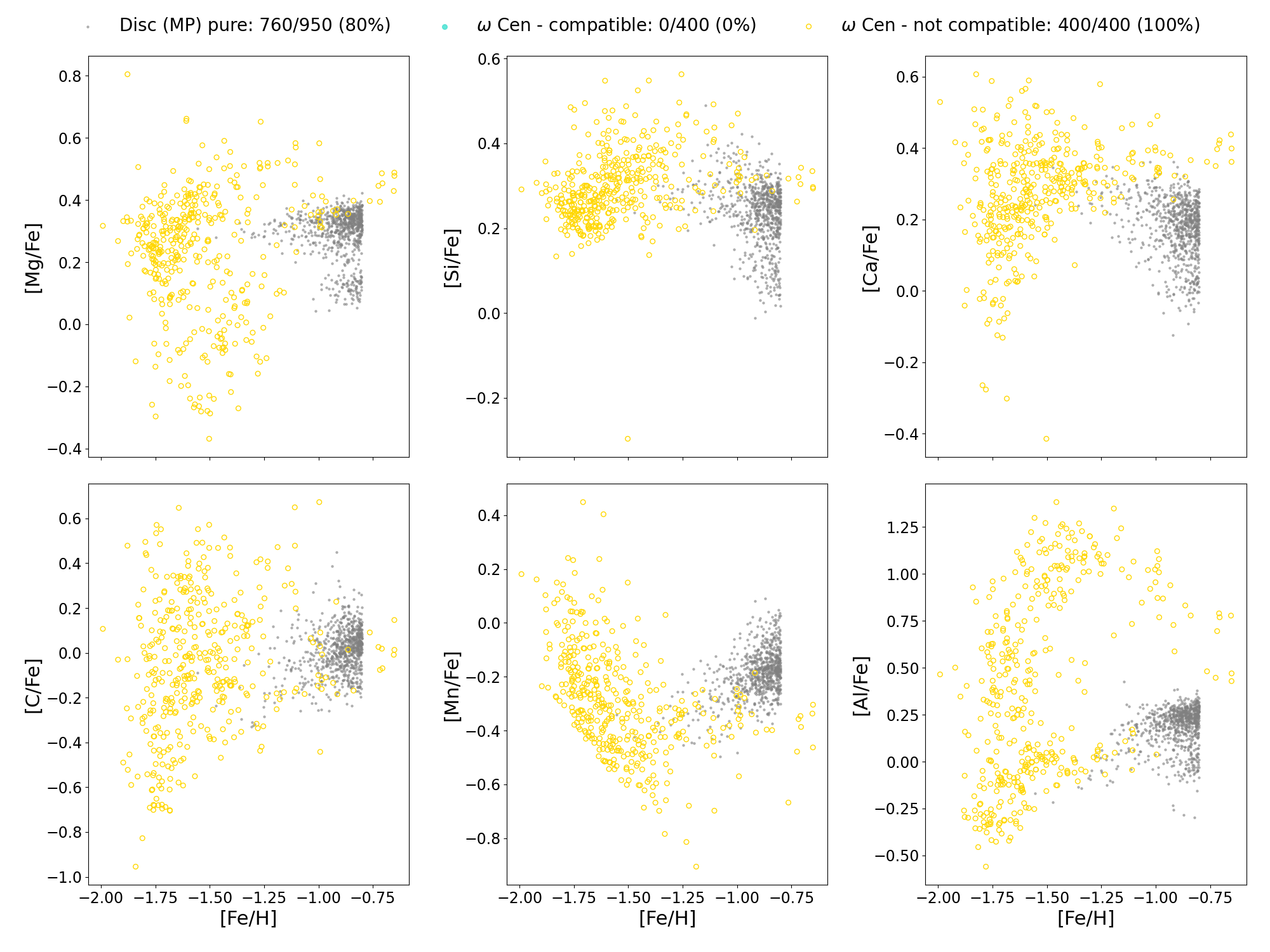}
    \caption{$\omega$ Cen's stars that are chemically compatible (threshold=20) with GSE (H) (upper panel) and with the metal-poor disc (lower panel) are shown in turquoise, while the ones that are not compatible in yellow. The relative percentages are reported in the legend (and have to be normalised to 80\% for absolute values). In grey we show the bulk (80\%) of GSE and of the metal-poor disc, in the upper and lower panels, respectively.}
    \label{fig:ocen_cut}
\end{figure*}

In this appendix, we show the detailed chemical comparison of each of the halo substructures with respect to GSE (H) and the metal-poor disc, in the upper and lower panels of Figures \ref{fig:heracles_cut}-\ref{fig:ocen_cut}, respectively. The extensive abundance distributions of the stars are reported in the following chemical spaces: [Mg/Fe], [Si/Fe], [Ca/Fe], [C/Fe], [Al/Fe], and [Mn/Fe] vs. [Fe/H] spaces.

In particular, in each figure we show in grey the stars of the bulk (80\%) of GSE (H) ("pure GSE", see Sec. \ref{sec:method}) in the upper panel and the ones of the bulk of the metal-poor disc in the lower panel, which are the samples to which we compare the other halo substructures. Then, for example, in the upper panel of Figure \ref{fig:heracles_cut}, we highlight in red the stars of Heracles that are identified as chemically compatible with the pure GSE (H) (by the GMM analysis with a threshold of 20), while in yellow the ones that are not. The total number of stars of the samples, the fraction of the ones shown (compatible or not), and the relative percentages are reported in the legend above the plot. We remind that these percentages are relative to the threshold assumed (20), so in this case the upper limit for the compatibility is 80\% (that is, they should be normalised to 80\% for absolute values). This threshold, however, allowed us to be very conservative and select the inner bulk of the GSE/disc distributions. The same is done for the comparison with the metal-poor disc ([Fe/H] $\le -0.8$) in the lower panel. 

Analogously, with the colour-coding of the main body of the article, in the following figures \ref{fig:sequoia_cut}, \ref{fig:thamnos_cut}, \ref{fig:helmi_cut}, \ref{fig:sgr_cut}, and \ref{fig:ocen_cut}, we show in green, brown, magenta, orange, and turquoise the stars of Sequoia (K), Thamnos, Helmi Stream (K), Sagittarius, and $\omega$ Centauri that are chemically compatible with GSE (H) and the metal-poor disc, respectively.

\section{Spatial and dynamical comparison of the cleaned Helmi Stream with Sagittarius} 
\label{appendix:helmisgr}

\begin{figure*}
    \centering
    \includegraphics[width=1\columnwidth]{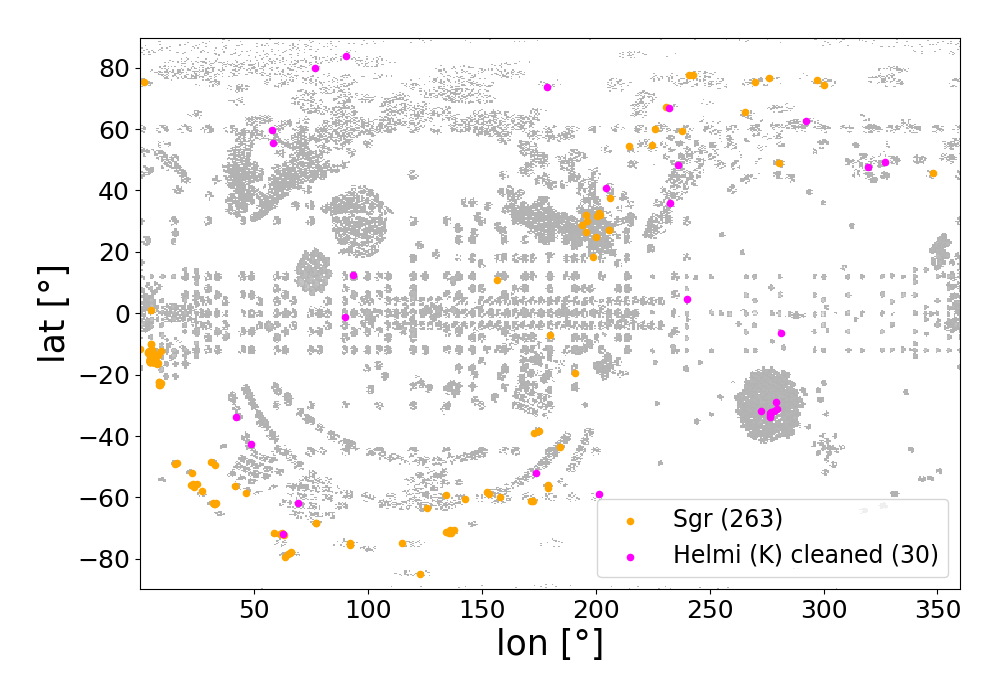}
    \includegraphics[width=1\columnwidth]{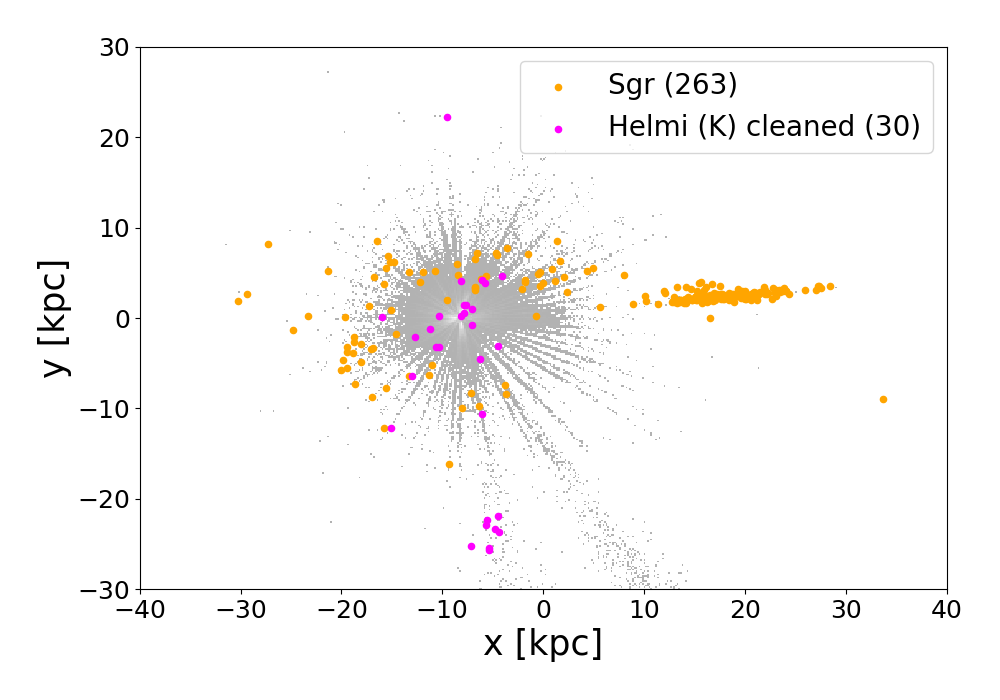}
    \includegraphics[width=0.98\columnwidth]{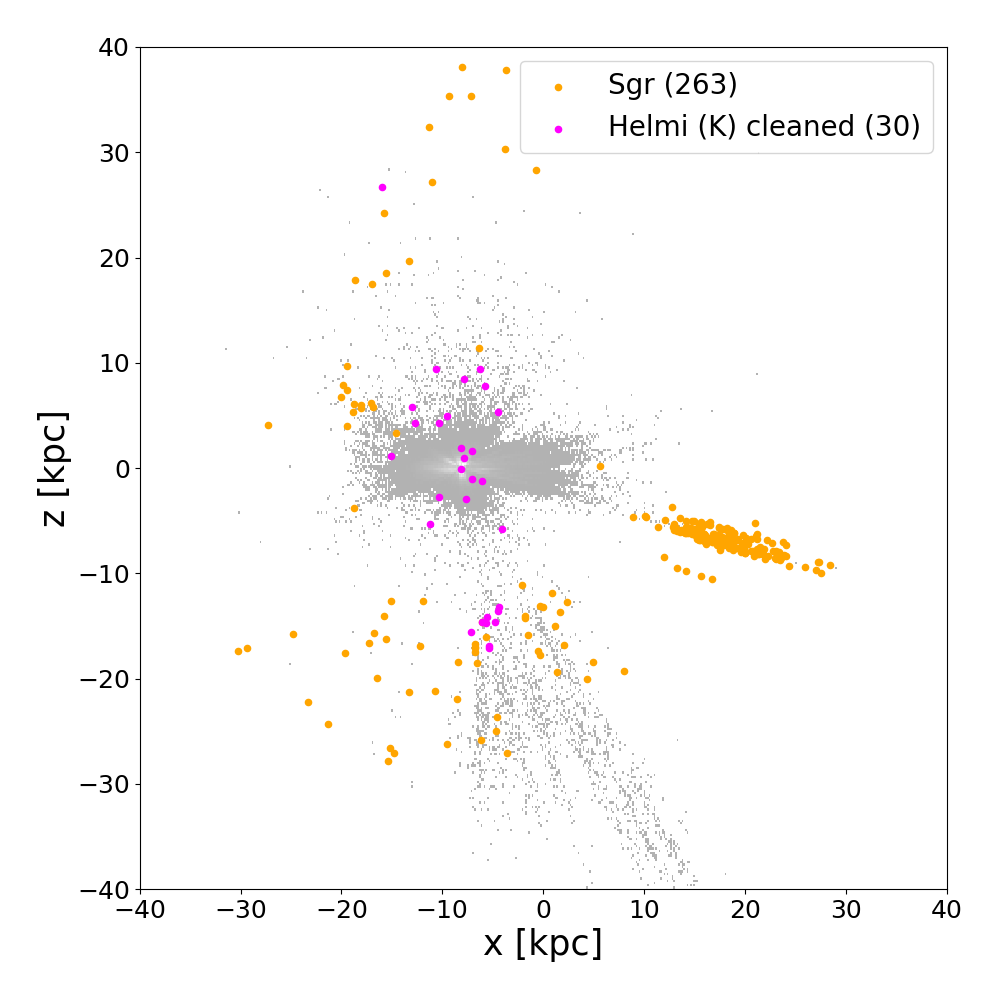}
    \includegraphics[width=0.98\columnwidth]{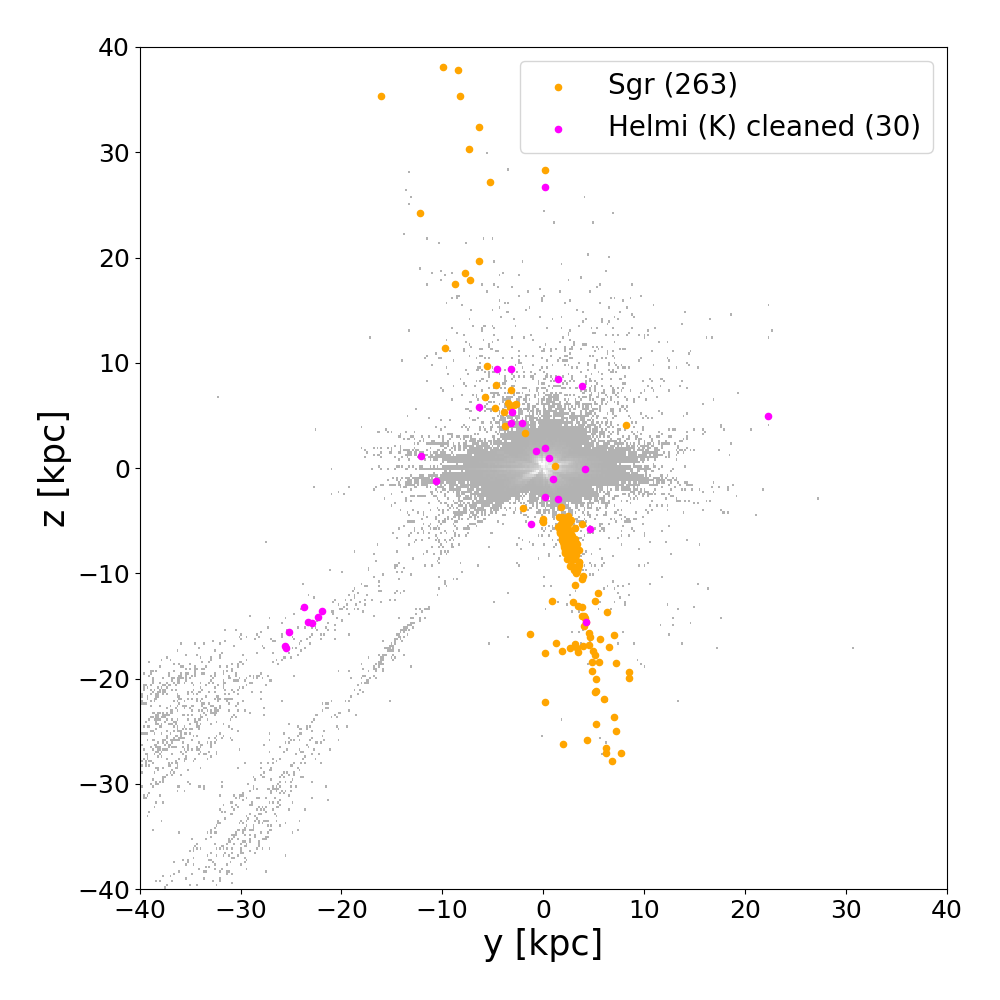}
    \includegraphics[width=0.98\columnwidth]{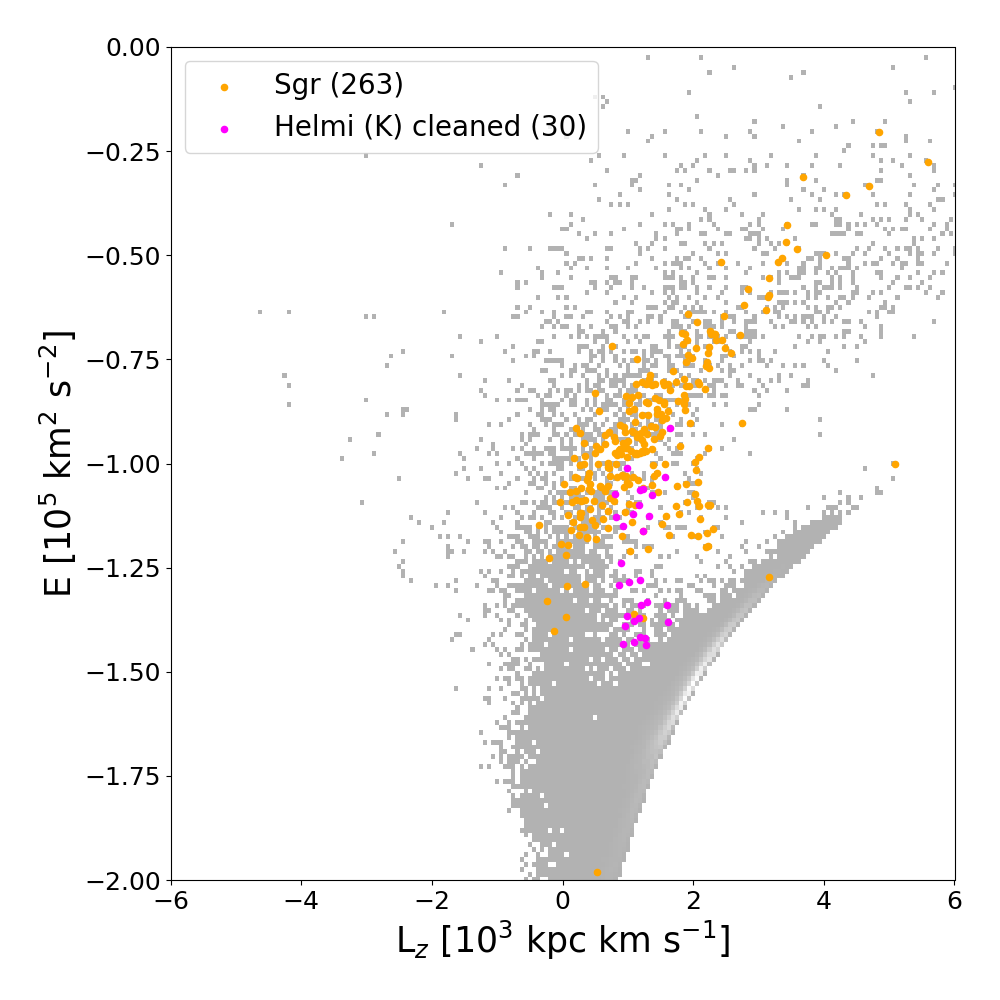}
    \includegraphics[width=0.98\columnwidth]{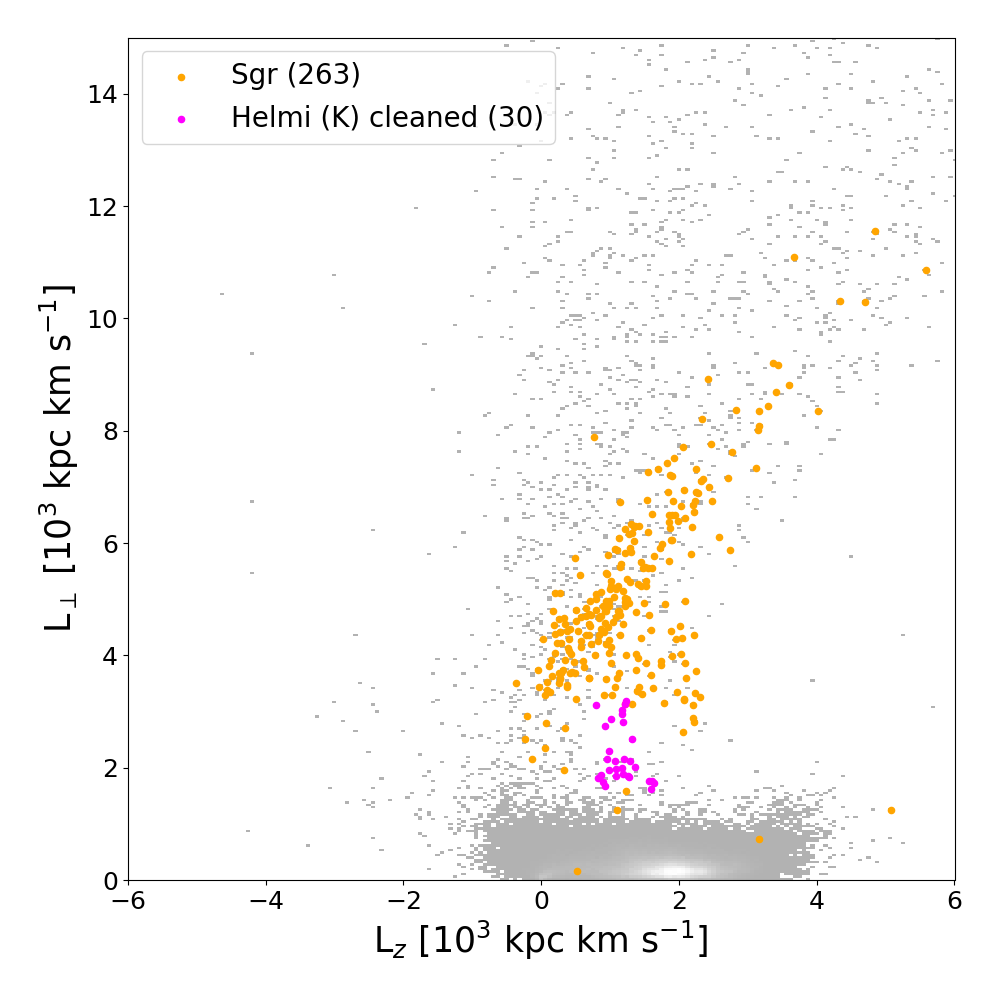}
    \caption{Spatial and dynamical distributions in the $lon-lat$, $x-y$, $x-z$, $y-z$, $E-L_z$, $L_\perp-L_z$ spaces of the cleaned Helmi Stream (in magenta), Sagittarius' stars (in orange) and the whole sample (in grey).}
    \label{fig:helmi-sgr-appendix}
\end{figure*}

A further comparison between the cleaned sample of the Helmi Stream and the Sagittarius is provided in this appendix. These are the stars in the Helmi Stream sample that are not compatible with either GSE (H) or with the metal-poor disc ([Fe/H]$\le-0.8$). 

The comparison between these two substructures was driven by the fact that they are close in the $E-L_z$ space: both systems are found on prograde orbits, with the Helmi Stream (in magenta) on lower energies but still partially overlapping with Sagittarius (in orange), as shown in the beginning by Figure \ref{fig:elz}. The same space is shown in the left panel of the bottom row in Figure \ref{fig:helmi-sgr-appendix} for the cleaned Helmi Stream sample (in magenta), all the Sagittarius stars (in orange) and the whole sample (in grey) for reference. The right panel of the same row displays - with the same colour legend - the $L_\perp-L_z$ space, where again Sagittarius lies on average on higher $L_\perp$ values, but populating the region occupied by the cleaned Helmi Stream as well. 

For a more complete characterisation of the cleaned Helmi Stream, we also show its spatial distribution in the $lon-lat$, $x-y$, $x-z$, and $y-z$ spaces in the two upper rows of Figure \ref{fig:helmi-sgr-appendix}.  
We can see that many of the cleaned Helmi Stream stars follow the Sagittarius Stream spatial distribution.

\end{appendix}

\end{document}